\documentclass[apj]{emulateapj}
\usepackage{textcomp}
\usepackage{epsfig, epstopdf}
\usepackage{amssymb,amsmath}

\newcommand{\CIV}{C~{\sc iv}}

\newcommand{\NV}{N~{\sc v}}

\newcommand{\FeXXV}{\hbox{Fe {\sc xxv}}}

\newcommand{\SiXIV}{\hbox{Si {\sc xiv}}}

\newcommand{\simgt}{\lower 2pt \hbox{$\, \buildrel {\scriptstyle >}\over {\scriptstyle\sim}\,$}}
\newcommand{\simlt}{\lower 2pt \hbox{$\, \buildrel {\scriptstyle <}\over {\scriptstyle\sim}\,$}}

\newcommand{\hs}{HS~0810+2554}
\newcommand{\rxj}{RX~J1131$-$1231}

\newcommand{\chandra}{{\emph{Chandra}}}

\newcommand{\xmm}{\emph{XMM-Newton}}

\slugcomment{Received 2015 Dec 27; Accepted 2016 Mar 16}
\shorttitle{The Wide-Angle Outflow of AGN HS~0810+2554}
\shortauthors{CHARTAS ET AL.}

\begin{document}

\def\sarc{$^{\prime\prime}\!\!.$}
\def\arcsec{$^{\prime\prime}$}
\def\beginrefer{\section*{References}%
\begin{quotation}\mbox{}\par}
\def\refer#1\par{{\setlength{\parindent}{-\leftmargin}\indent#1\par}}
\def\endrefer{\end{quotation}}

\title{ The Wide-Angle Outflow of the Lensed $\MakeLowercase{z}$  = 1.51 AGN HS~0810+2554}

\author{G. Chartas\altaffilmark{1,2}, M. Cappi\altaffilmark{3}, F. Hamann\altaffilmark{4,5}, M. Eracleous\altaffilmark{6}, S.~Strickland\altaffilmark{1}, M. Giustini\altaffilmark{7}, and T.~Misawa\altaffilmark{8} }

\altaffiltext{1}{Department of Physics and Astronomy, College of Charleston, Charleston, SC, 29424, USA, chartasg@cofc.edu}
\altaffiltext{2}{Department of Physics and Astronomy, University of South Carolina, Columbia, SC, 29208}
\altaffiltext{3}{INAF-Istituto di Astrofisica Spaziale e Fisica cosmica di Bologna, via Gobetti 101, I-40129, Bologna, Italy}
\altaffiltext{4} {Department of Astronomy, University of Florida, 211 Bryant Space Science Center, Gainesville, FL 32611-2055}
\altaffiltext{5} {Department of Physics \& Astronomy, University of California, Riverside, 900 University Avenue, Riverside, CA,92521, USA} 
\altaffiltext{6}{Department of Astronomy \& Astrophysics, Pennsylvania State University, University Park, PA 16802}
\altaffiltext{7}{SRON - Netherlands Institute for Space Research, Sorbonnelaan 2, 3584 CA Utrecht, Netherlands}
\altaffiltext{8}{School of General Education, Shinshu University, 3-1-1 Asahi, Matsumoto, Nagano 390-8621, Japan}

\begin{abstract}

\noindent 
We present results from X-ray observations of the gravitationally lensed $z~=~1.51$ AGN \hs\ performed with the {\sl Chandra X-ray Observatory} and {\sl XMM-Newton}. 
Blueshifted absorption lines are detected in both observations at rest-frame energies ranging between $\sim$1--12~keV at $\simgt$ 99\% confidence.
The inferred velocities of the outflowing components range between $\sim$0.1$c$ and $\sim$0.4$c$.

A strong emission line at $\sim$6.8~keV accompanied by a significant absorption line at $\sim$7.8~keV is also detected in the {\sl Chandra} observation. 
The presence of these lines is a characteristic feature of a P-Cygni profile supporting the presence of an expanding outflowing highly ionized iron absorber in this quasar.
Modeling of the P-Cygni profile constrains the covering factor of the wind to be $\simgt$~0.6, assuming disk shielding. 
A disk-reflection component is detected in the {\sl XMM-Newton} observation accompanied by blueshifted absorption lines. 
The \xmm\ observation constrains the inclination angle to be $<$~45$^{\circ}$ at $90\%$ confidence, assuming the hard excess is due to blurred reflection from the accretion disk. 

The detection of an ultrafast and wide-angle wind 
in an AGN with intrinsic narrow absorption lines (NALs) 
would suggest that quasar winds may couple efficiently with the intergalactic medium and provide significant feedback if ubiquitous in all NAL and BAL quasars.

We estimate the mass-outflow rate of the absorbers to lie in the range of 1.5 and 3.4 M$_{\odot}$~yr$^{-1}$ for the two observations.
We find the fraction of kinetic to electromagnetic luminosity released by \hs\ is large ($\epsilon_{\rm k}$~=~9$_{-6}^{+8}$) suggesting that magnetic driving is likely a significant contributor to the acceleration of this outflow. 
\end{abstract}


\section{INTRODUCTION}

Our X-ray survey of Narrow Absorption Line (NAL; FWHM $ $\simlt$ $500~km~s$^{-1}$) AGN with outflows  (Chartas et al., 2009a)  has revealed a remarkable X-ray bright gravitationally lensed object: HS~0810+2554. It was detected in a 5~ks $\it Chandra$ observation with a  2--10~keV observed  flux of {8$\times$10$^{-13}$~erg~s$^{-1}$~cm$^{-2}$} 
and its spectrum shows hints of blueshifted high energy broad absorption lines (BALs) implying the presence of a massive and ultrafast X-ray absorbing wind (Chartas et al. 2014).
The VLT/UVES spectrum of \hs\ shows blueshifted \CIV\ and \NV\ doublets that are likely intrinsic to the quasar outflow based on a partial covering analysis (see Chartas et al. 2014 and references therein). The observed wavelengths of the \CIV\ doublet ($\lambda\lambda$ 1548.20 $\rm \AA$, 1550.77 $\rm \AA$) imply that the 
UV absorber is outflowing with a speed of $v_{\rm C IV}$ $\sim$ 19,400 km~s$^{-1}$.  
The only other quasar detected to date with similar blueshifted X-ray absorption features
and a narrow absorption line (NAL) outflow is the unlensed $z$=2.74 NAL quasar HS~1700+6414 (Lanzuisi et al. 2012)
which, however, has a $\sim$10 times lower 2--10 keV  flux of {3--9}$\times$10$^{-14}$ erg~s$^{-1}$~cm$^{-2}$.

Blueshifted X-ray broad absorption lines have been detected in several high-$z$ quasars (i.e., APM~08279+5255, Chartas et al. 2002; PG~1115+080, Chartas et al. 2003;  and HS~1700+6416, Lanzuisi et al. 2012). 
Our spectral analysis of the 5~ks $\it Chandra$ observation revealed a possible near-relativistic outflow of X-ray absorbing material also in \hs\ (Chartas et al. 2014). Because of the low signal-to-noise ratio of the 5~ks Chandra observation only the combined spectrum of all images was analyzed. The main result from that analysis was the detection of absorption lines at rest-frame energies of $\sim$7.7~keV and $\sim$10~keV at the 97\% significance level implying the presence of two outflowing components at velocities of $\sim$0.13$c$ and  $\sim$0.41$c$ respectively.

We recently re-observed \hs\ for $\sim$98~ks with {\sl Chandra} and for ${\sim}$46.4~ks with {\sl XMM-Newton.} The goals of the follow-up observations of \hs\ were to:
(a) confirm the ultrafast outflow of HS 0810+2554 at different epochs and in individual images, (b) constrain the properties of the outflowing material in the higher S/N observations, (c) study the variability of the X-ray outflow, and, (d) determine the contribution of a galaxy group to the lensing of HS 0810+2554. In this paper we present results from these observations of \hs\ and address the science goals that we set out to meet.
 
In $\S$2 we present the spectral and spatial analysis of \hs\ and a nearby galaxy group that contributes to the lensing, in $\S$3 we discuss the properties of the outflow and show
how our estimated values of the mass-outflow rate, wind efficiency and covering factor imply that the quasar wind in \hs\ provides significant feedback through a 
massive, energetic, and wide-angle outflow, and in $\S$4 we present a summary of our conclusions. Throughout this paper we adopt a flat $\Lambda$ cosmology with $H_{0}$ = 67~km~s$^{-1}$~Mpc$^{-1}$,  $\Omega_{\rm \Lambda}$~=~0.69, and  $\Omega_{\rm M}$ = 0.31, based on the Planck 2015 results (Planck Collaboration et al. 2015).

\section{X-RAY OBSERVATION AND DATA ANALYSIS}
\hs\ was observed with the Advanced CCD Imaging Spectrometer (ACIS; Garmire et al. 2003) on board 
the {\sl Chandra X-ray Observatory} (hereafter {\sl Chandra}) on 
2002 January 30 and 2013 December 16 with effective exposure times of  4,894~sec and 97,730~sec, respectively.
\hs\ was also observed with {\sl XMM-Newton} (Jansen et al. 2001) on October 4, 2014 with an effective exposure of 46,400~sec.
The analysis of the 2002 observation was presented in Chartas et al. 2014. 
Here we describe the analysis of the 2013 and 2014 observations of \hs.

For the reduction of the {\sl Chandra} observations we used 
the CIAO 4.6 software with CALDB version 4.5.9 provided by the {\sl Chandra X-ray Center} (CXC).
We used standard CXC threads to screen the data for status, grade, and time intervals of acceptable aspect solution and background levels. 
To improve the spatial resolution we employed the sub-pixel resolution technique developed by Li et al. (2004) and incorporated via the
Energy-Dependent Subpixel Event Repositioning (EDSER) algorithm into the tool acis\_process\_events of CIAO 4.5.
The pointing of the telescope placed \hs\ on the back-illuminated S3 chip of ACIS.

For the reduction of the \xmm\ observations we filtered the pn (Str{\" u}der et al. 2001) and MOS (Turner et al. 2001) data by selecting events corresponding to instrument \verb+PATTERNS+ in the 0--4 (single and double pixel events) and 0--12 (up to quadruple pixel events) ranges, respectively. Several moderate-amplitude background flares were present
during the last $\sim$7~ks of the \xmm\ observation of \hs. The pn and MOS data were filtered to exclude times when the full-field-of-view count rates exceeded 30~cnts~s$^{-1}$ and 4~cnts~s$^{-1}$, respectively.  

To test the sensitivity of our results to background flares we also extracted pn and MOS spectra with more conservative values of the full-field-of-view count rates of 20~cnts~s$^{-1}$ and 3~cnts~s$^{-1}$, respectively.  To test for sensitivity to background non-uniformity we also tried different background extraction regions. We did not find any differences in the spectral shapes and features using the more conservative threshold cuts or selecting different background extraction regions.

The pn and MOS spectra were fited with a variety of models employing \verb+XSPEC+ version 12 (Arnaud 1996).  The energy ranges used for fitting the pn and MOS 
data were 0.3--11~keV and 0.4--10~keV, respectively. We performed spectral fits to the pn spectra alone and to the pn and MOS data simultaneously.
Both approaches resulted in values for the fitted parameters that were consistent within the errors, however, the fits to the higher quality pn data alone 
yielded lower reduced $\chi^{2}$ values compared to the combined fits. We therefore consider the results from the fits to the pn data alone more reliable especially for characterizing the properties of the X-ray absorption features.

The \chandra\ spectra of the individual images of \hs\ were also fitted with a variety of models employing \verb+XSPEC+.
For all spectral models of \hs\ we included Galactic absorption due to neutral gas (Dickey \& Lockman 1990) with a column density of 
 $N_{\rm H}$= 3.94 $\times$ 10$^{20}$~cm$^{-2}$.

\subsection{Spatial Analysis of \hs}

We perform a spatial analysis of the {\sl Chandra} observation of \hs\ by fitting  the 
lensed images (see Figure 1) simultaneously using the relative astrometry derived from {\sl HST} images 
and \verb+MARX+ models (Wise et al. 1997) of the PSF.
The details of this analysis are similar to the procedure described in Chartas et al. 2014.
Table 1 lists the observation date, the observational identification number, the exposure time, 
and the number of counts (in the 0.2--10~keV band) of each image corrected for PSF effects.
For comparison we also include the results from the 2002 observation of \hs.
The total 0.2-10~keV count-rate has decreased by a factor of $\sim$ 2.5 between the 2003 and 2013 epochs.
\xmm\ does not resolve the lensed images of \hs\ and in Table 1 we list the total background-subtracted EPIC-pn counts from all images. 

Figure 1 shows a binned image and the best-fit PSF model of the 2013  {\sl Chandra} observation of \hs.
In Figure 2 we show the deconvolved image. For the deconvolution we applied the 
Richardson-Lucy algorithm (Richardson 1972; Lucy 1974)
and supplied a point spread function (PSF) created by the simulation tool \verb+MARX+ (Wise
et al. 1997). Images A and B are not resolved, while image C is resolved in the deconvolution. Image D is likely too faint to
be reconstructed in the deconvolution.

The multi-wavelength flux ratios of \hs\ are presented in Table 2. Image C is significantly brighter in the X-ray band in the 2013 epoch compared to the 2002 epoch and its X-ray flux ratio differs significantly from the optical. Images C and D are easily resolved by {\sl Chandra} and {\sl HST} and their flux ratios 
are less affected by any possible systematic effects due to PSF modeling.  
One possible explanation of this flux anomaly of image C is strong magnification due to microlensing
of this image during the 2013 observation. Microlensing will in general distort the X-ray spectrum of the affected image. We therefore fit the {\sl Chandra} spectra of images (A+B)
and C separately (see $\S$2.2.1).

\subsection{Spectral Reduction Procedures of \hs}

For the spectral analysis of the {\sl Chandra} observations we extracted spectra for the combined images A+B and for the spatially resolved image C. Image D was too faint to extract a spectrum. The extracted regions are shown in Figure 1. The background were determined by extracting events within an annulus centered on the mean location of the images with inner and outer radii of 10~arcsec and 50~arcsec, respectively. The extracted {\sl Chandra} spectra were grouped to obtain a minimum of 20 counts in each energy bin, above the minimum required number of counts per bin for $\chi^2$ to be statistically valid (e.g., Cash 1979; Bevington \& Robinson 2003).

For the spectral analysis of the \xmm\ observation we extracted spectra for the combined images A+B+C+D. The extracted spectra from the pn and MOS 
were grouped to obtain a minimum of 40 counts in each energy bin, allowing use of $ \chi^{2}$ statistics.
We tested the sensitivity of our results to the grouping factor by performing the analysis with the grouping ranging between 20 counts per bin and 100 counts per bin.
The best-fit parameters were found to be similar in all cases.  
We selected a grouping of 40 counts per bin because this provided tight constraints on the fitted parameters 
with the bin size being comparable to or smaller than the energy resolution of the instrument in regions of the spectrum where we expect to find narrow absorption or emission lines.  
Background spectra for the pn and MOS detectors were extracted from source-free regions near \hs.

We performed fits to the spectra of \hs\ with a variety of models of increasing complexity. The fit residuals show significant absorption and/or emission at observed-frame energies of 1--8~keV. To illustrate the presence of these features we fit the {\sl Chandra} spectra of combined images A+B and resolved image C (see Figure 3)
and the \xmm\ spectra of combined images A+B+C+D (see Figure 4) in the observed-frame 0.35--10~keV band with a power-law model
(modified by Galactic absorption) and plot the best-fit model and the fit residuals. 

We proceed in fitting the following models to the data (see Tables 3, 4, and 5) guided by the shape and location of  identified 
absorption and/or emission residuals:  

\noindent
1) power-law modified by neutral intrinsic absorption at the source.
This and all subsequent models include Galactic absorption due to neutral gas (Dickey \& Lockman 1990) with a column density of 
 $N_{\rm H}$= 3.94 $\times$ 10$^{20}$~cm$^{-2}$.

\noindent
2) power-law modified by neutral intrinsic absorption and a number of absorption and/or emission lines.
The {\sl Chandra} spectra of images (A+B) and C differ significantly due to microlensing (see $\S$ 2.2.1).
Guided by the residual features to fits using model 1 (see Figure 1) the {\sl Chandra} spectrum of image A+B was next fitted with a model that included three absorption lines and one emission line and the {\sl Chandra} spectrum of image C was fitted with a model that included three emission lines. The {\sl Chandra} spectrum of image (A+B) and the XMM-Newton spectrum of (A+B+C+D) differ significantly due to the fact that {\sl XMM-Newton} does not resolve the images and due to possible spectral variability between the observations. Guided by the residual features to the fits using model 1 (see Figure 4) the {\sl XMM-Newton} spectrum was next fitted with a model that included two absorption lines and neutral Compton reflection with self-consistent Fe and Ni lines (PEXMON). In model 2 the absorption lines are fit with Gaussians. In the next model the Gaussian absorption lines are replaced with a more realistic model of outflowing intrinsic ionized absorption.

\noindent
3) power-law modified by neutral intrinsic absorption, outflowing intrinsic ionized absorption 
and a number of emission lines. Specifically, the {\it Chandra} spectrum of image A+B was fitted with a model that
included one outflowing ionized absorber and one emission line, the {\it Chandra} spectrum of image C was fitted with a model that
included one outflowing ionized absorber and three emission lines, and the {\it XMM-Newton} spectrum was fitted with a model that
included two outflowing ionized absorbers and neutral Compton reflection with self-consistent Fe and Ni lines (PEXMON).

\noindent
4) power-law modified by neutral intrinsic absorption, an outflowing ionized absorber, and scattering by the outflowing wind.
This model was fitted to the {\it Chandra} spectrum of image A+B.

\noindent
5) power-law modified by neutral intrinsic absorption, a narrow emission line near the Fe~K$\alpha$ line, an outflowing ionized absorber, and scattering by the outflowing wind. This model was fitted to the {\it Chandra} spectrum of image A+B.

\noindent
6) power-law modified by neutral intrinsic absorption, an outflowing ionized absorber, scattering by the outflowing wind, and neutral Compton reflection from distant parts of the accretion disk.
This model was fitted to the {\it XMM-Newton} spectrum of \hs.

\noindent
7) power-law modified by neutral intrinsic absorption, an outflowing ionized absorber, scattering by the outflowing wind, and retativistically blurred X-ray reflection from parts of the accretion disk near the black hole. This model was fitted to the {\it XMM-Newton} spectrum of \hs.

Models 4 and 5 are only used to fit the {\it Chandra} spectrum of image A+B that shows a noticeable P-Cygni profile.
Models 6 and 7 are only used to fit the {\it XMM-Newton} spectrum that shows a noticeable hard-X-ray excess above 15~keV in the rest-frame.

For the outflowing ionized absorber with no scattering we used the \verb+XSTAR+ photoionization model \verb+warmabs+ 
and for an outflowing ionized absorber with scattering off the outflowing absorber we used the \verb+XSTAR+  model  \verb+windabs+ (Kallman et al. 1996; Kallman \& Bautista 2001). For improved accuracy and flexibility we have selected to use the analytic \verb+XSTAR+ versions of the \verb+warmabs+ and \verb+windabs+ models instead of the \verb+XSTAR+ table models. 

We model the velocity broadening of the absorption lines by introducing in the
\verb+XSTAR+  models large turbulent velocities.
We performed several fits where we allowed the turbulence velocity to vary and found the best-fit values.
Because of the low to moderate S/N of the {\sl Chandra} and {\it XMM-Newton} spectra of \hs, the
turbulent velocities, are not well constrained.
For the error analysis of the remaining variables in spectral fits that used the \verb+XSTAR+  model
we therefore froze the turbulent velocities and list these values in Tables 3, 4, and 5.

For the Compton reflection we used the  \verb+PEXMON+ model 
(Nandra et al. 2007) that incorporates neutral Compton reflection with self-consistent Fe and Ni lines.
For the relativistically blurred X-ray reflection we used the RELXILL model (Garcia et al. 2014). 
The results from fitting these models to the \chandra\ and \xmm\ spectra of \hs\ are presented in  
Tables 3, 4, and 5.

\subsubsection{Spectral Analysis of {\sl Chandra} Observation of \hs}

We begin with a description of the fits to the {\it Chandra} spectrum of image (A+B) of \hs.
The intrinsic neutral absorption in image A+B is found to be relatively low with $N_{\rm H}$ = 0.39$^{+0.15}_{-0.13}$ $\times$ 10$^{22}$~cm$^{-2}$ 
(model 3 of Table 3).
In Figure 5 we show the 68\%, 90\% and 99\% $\chi^{2}$ confidence contours of the 
intrinsic hydrogen column density, $N_{\rm H}$ versus the photon index.
We consider our constraints on the intrinsic $N_{\rm H}$ from the fits to the  {\it Chandra} spectrum of A+B 
more reliable than the fits to the spectrum of image C since image C is likely microlensed 
based on the X-ray flux ratios.

The {\it Chandra} spectra of images A+B and C differ significantly and thus were fitted separately.
A broad emission line at 6.8$^{+0.2}_{-0.2}$ keV is detected in the spectrum of image A+B and 
significant absorption lines at energies at 2.2$^{+0.1}_{-0.1}$ keV, 7.8$^{+0.5}_{-0.5}$ keV, and 13.1$^{+1.6}_{-1.5}$ keV 
(all energies expressed in the rest-frame).
The inclusion of three Gaussian absorption lines near the absorption features and one Gaussian emission line in model 2 of Table 3 resulted in a significant improvement of the fit compared to the fit using model 1 ($\Delta\chi^{2}$ =  29.8 ${\Delta}$${\sl dof}$ = 12, where ${\sl dof}$ are the degrees of freedom).
$\chi^{2}$ confidence contours of the normalizations of the absorption lines versus their rest frame energies based on fits using model 2 of Table 3 
are presented in Figure 6.
The absorption line at 7.8~keV is detected at $>$ 99\% confidence while the lines at 2.2~keV and 13.1~keV are marginally detected at $>$~95\% confidence.
The energy of the emission line at $\sim$ 6.8 keV is consistent with Fe~XXV and/or Fe~XXVI.
The presence of this emission line at $\sim$ at 6.8~keV and an absorption line at $\sim$ 7.8 keV is a characteristic feature of a P-Cygni profile
supporting the presence of an expanding outflowing highly ionized absorber from quasar HS 0810+2554.
$\chi^{2}$ confidence contours of the normalizations of the emission and absorption line of the P-Cygni profile are presented in Figure 7.
The P-Cygni lines are detected at $>$ 99\% confidence.
The ratio of the normalizations of the emission to absorption lines of the P-Cygni profile of image A+B is $R_{\rm P}$~=~$N_{\rm emis}/N_{\rm abs}$ $\sim$ 0.75.
We note that the ratio measured for the wide-angle outflow of the nearby quasar PDS~456 over five different epochs lies in the range of $R_{\rm P}$ = 0.19 --1.67 with an average value of $ \left<  R_{\rm P} \right>$ $\sim$ 0.7 (Nardini et al. 2015).

We propose as a possible explanation for the difference between the spectrum of fainter image C and the combined spectrum of images A+B a microlensing event occurring in image C. This interpretation is consistent with the anomalous X-ray flux ratio of image C (see Table 4).
Microlensing may lead to energy shifts of the iron emission lines due to general and relativistic effects as found in the
lensed quasar \rxj (Chartas et al. 2016). The energies of the best-fit emission lines of image C are presented in Table 3.
We note that the emission lines in image C are only detected at the $\sim$ 90\% confidence level.
We fit the {\it Chandra} spectrum of (A+B)  with a model that includes an outflowing photoionized absorber and one emission line (model 3 of Table 3).
This model can account for the absorption features at around 2.2~keV and 7.8~keV as originating from highly ionized Si and Fe absorbers, respectively. 

To model the P-Cygni profile (emission and absorptions lines at $\sim$ 6.8 keV and $\sim$ 7.8 keV, respectively) we used the \verb+XSTAR+  model  \verb+windabs+ (model 4 of Table 4). The fit with model 4 is acceptable in a statistical sense, however, a residual emission feature remains near the energy of the Fe K$\alpha$ line. To account for this emission feature we add a narrow Gaussian line (model 5 of Table 4).
A narrow Fe line component is typically detected in the spectra of AGN along with the relativistically blurred Fe Ka line. The narrow Fe line component is thought to originate in distant, cold matter such as a molecular torus or an optical broad line region (e.g. Krolik et al. 1994; Yaqoob \& Padmanabhan 2004; Zhou \& Wang 2005). 
Another plausible process that could produce the residual emission feature near 6.4~keV is 
 scattering from a low velocity component of the wind.

In Figure 8 we show the {\it Chandra} spectrum of A+B overplotted with the best-fit model (model 5 of Table 4). Figure 8 also shows the unfolded best-fit model that includes the P-Cygni profile, a characteristic spectral signature of an outflowing wind. 
We consider model 5 of Table 4 as our most realistic model that provides a statistically acceptable fit to the {\it Chandra} spectrum of image (A+B) of \hs.
We base any following estimates of the outflow properties for this epoch on the best-fit values of model 5 of Table 4.
The $\chi^{2}$ confidence contours of the covering factor of the wind based on models 4 and 5 of Table 4 are shown in Figure 9. The best-fit values of the covering factor are 
$f_{\rm c}$~=~0.57$_{-0.18}^{+0.18}$ (68$\%$ confidence)  assuming model 4 and  $f_{\rm c}$~=~0.6$_{-0.3}^{+0.2}$ (68$\%$ confidence) for model 5.

Model 5 of Table 4 can account for the absorption features at around 2.2~keV and 7.8~keV as originating from highly ionized Si and Fe absorbers, respectively,  outflowing with a common velocity of 0.1$^{+0.01}_{-0.01}c$.
The outflow is highly ionized with an ionization parameter \footnote{Throughout this paper we
adopt the definition of the ionization parameter of Tarter et al. (1969) given by
$\xi=\frac{L_{\rm ion}}{n_H r^2}=\frac{4 \pi}{n_H}
\int_{1Rdy}^{1000Rdy}F_{\nu}d\nu$, where $n_H$ is the hydrogen number
density, and $r$ is the source-cloud separation.} 
of $\log(\xi/\rm erg~cm~s^{-1})$ = 3.15$^{+0.13}_{-0.21}$ and a column density of 
$N_{\rm H,out}$ = 2.9$^{+1.6}_{-0.4}$ $\times$ 10$^{23}$~cm$^{-2}$.

\subsubsection{Spectral Analysis of {\sl XMM-Newton} Observation of \hs}

We next proceed in the analysis of the {\it XMM-Newton} spectrum of images A+B+C+D of \hs. Our fit to the {\it XMM-Newton} spectrum with a simple power-law with Galactic absorption shows residual absorption features and a hard excess above $\sim$ 15 keV in the rest-frame.
We fit the spectrum of {\it XMM-Newton} of \hs\ with models 1$-$3 with the best-fit parameters listed in Table 3. Significant absorption lines are detected at energies of  7.9$^{+0.2}_{-0.3}$ keV and  10.9$^{+0.7}_{-0.6}$ keV (all energies expressed in the rest-frame).
The inclusion of two Gaussian absorption lines near the absorption features in model 2 of Table 3 and neutral Compton reflection with self-consistent Fe and Ni lines resulted in a significant improvement of the fit compared to the fit using model 1 ($\Delta\chi^{2}$ =  44.5, ${\Delta}$${\sl dof}$ = 9). $\chi^{2}$ confidence contours of the normalizations of the absorption lines versus their rest frame energies based on fits using model 2 of Table 3 are presented in Figure 10.
We continue by fitting the {\it XMM-Newton} spectrum of \hs\ with a model that includes two outflowing photoionized absorbers and reflection (model 3 of Table 3). 

We next included scattering from the outflow (\verb+windabs+) to account for a possible weak emission feature near rest-frame $\sim$ 7.5 keV  (model 6 of Table 5). 
We finally considered a model (model 7 of Table 5) that describes the hard excess as relativistically blurred X-ray reflection in place of the neutral Compton reflection considered in model 6. 
The fit with model 7 implies the presence of two outflowing components with ionization parameters of $\log(\xi/\rm erg~cm~s^{-1})$ = 3.3$^{+0.3}_{-0.1}$  and $\log(\xi/\rm erg~cm~s^{-1})$ = 3.2$^{+0.2}_{-0.1}$,
column densities of $N_{\rm H}$ = 3.4$^{+1.9}_{-2.0}$ $\times$ 10$^{23}$~cm$^{-2}$ and $N_{\rm H}$ = 2.9$^{+2.0}_{-1.6}$ $\times$ 10$^{23}$~cm$^{-2}$, 
and outflow velocities of 0.12$^{+0.02}_{-0.01}$$c$ and 0.41$^{+0.07}_{-0.04}$$c$, respectively.  
In Figure 11 we show that the fit using model 7 of Table 5 constrains the inclination angle of the accretion disk to be less than 45$^{\circ}$ (90\% confidence).

The main difference between models 6 and 7 is that model 6 assumes distant reflection whereas model 7 assumes a relativistic, hence nearby reflector. 
Fits to the {\sl XMM-Newton} spectrum of \hs\ using models 6 and 7 are acceptable with similar (within the uncertainties) best-fit parameters for the properties of the outflowing absorbers, the reflection fraction and the inclination angle of the disk (model 7 provides tighter constraints on the inclination angle). 
We therefore consider models 6 and 7 of Table 5 as our most realistic models that provide statistically acceptable fits to the {\it XMM-Newton} spectrum of image (A+B+C+D) of \hs. 
We base any following estimates of the outflow properties for this epoch on the best-fit values of model 7 of Table 5, however, we note that the results are similar had we chosen to use the best-fit results from model 6.

We note that the spectral fit to {\sl XMM-Newton} spectrum requires two ionized absorbers and the fit to the {\sl Chandra} spectrum only one.
As indicated in Figure 6, the absorption line at $\sim$~13 keV is only marginally  detected in the {\sl Chandra} spectrum and no useful constraints can be obtained by including a second ionized absorber to model this line in the {\sl Chandra} spectrum. {\sl XMM-Newton} is more efficient than {\sl Chandra} at higher energies, which may explain why the second high-energy absorption line is detected at a higher confidence level in the {\sl XMM-Newton} spectrum.

\subsection{A Group of Galaxies Near \hs}

In Figure~12 we show the 100~ks {\it Chandra} image of  \hs\ and a nearby group of galaxies that contributes to the lensing of this quasar. The image was binned with a bin size of 0.1 arcsec and adaptively smoothed with the CSMOOTH tool developed by Ebeling et al. (2000).The 100~ks {\it Chandra} observation compared to the previous 5~ks one provides better constraints of the mass profile and the temperature of this galaxy group.These mass estimates were used to evaluate the convergence parameter ${\kappa}(x)$, 

\begin{displaymath}
{\kappa(x) = {{\Sigma(x)}\over{\Sigma_{cr}}}} \,
\end{displaymath}
where ${\Sigma(x)}$ is the surface mass density of the galaxy group
as a function of the cylindrical radius $x$ (e.g., Chartas et al. 1998) and
$\Sigma_{cr}$ is the critical surface mass density 
(see, e.g., Schneider, Ehlers \& Falco 1992). 
The distortion to the source produced by convergence is isotropic. An external perturber will also produce a distortion referred to as shear ($\gamma$) that stretches the intrinsic shape of the source in a preferred  direction. The shear from the galaxy group can be expressed as 
${\gamma} = {{\kappa}\over{({1 +{\beta_{rd}}^2})^{3/2}}}$ 
where, ${\beta_{rd}} = r_{c}/d_{c}$, $r_{c}$
is the galaxy group core radius and $d_{c}$ is the distance from the
group center to the center of \hs\ (see, Kochanek, 1991).

The spatial and spectral analysis of the galaxy group follows steps similar to those presented in Chartas et al. (2014).
Our revised fit to the X-ray brightness profile of the galaxy group using a $\beta$ model 
indicates best-fit values for its ellipticity of ${\epsilon} = $ 0.08$\pm$0.04,  $\beta$ =  0.38 $\pm$ 0.01, and for the core radius of the group of r$_{0}$ = 0{\sarc}24$_{-0.05}^{+0.05}$ (0.37~kpc).
We extract the {\it Chandra} and {\it XMM-Newton} spectra of the galaxy group from a 50 arcsec radius circle centered on the X-ray group center. The  spectra were jointly fit with a model consisting of emission from hot diffuse gas based on the XSPEC model {\it mekal} (Mewe, et al. 1985; Kaastra 1992; Liedahl, et al. 1995) modified by Galactic absorption. 
We obtain best-fit values for the redshift, temperature and metal abundances
of  $z$ = 0.082$_{-0.009}^{+0.007}$, $T_{\rm e}$~=~0.81$_{-0.03}^{+0.02}$~keV and $A$ = 0.53$^{+0.08}_{-0.11}$ solar, respectively (all errors are at the 90\% confidence level). 
The {\it Chandra} and {\it XMM-Newton} spectra of the galaxy group overplotted with our best-fit model are shown in Figure~13. 
The convergence and external shear due to the group at the lens position based on our analysis of the 100~ks {\sl Chandra} observation 
are found to be $\kappa$ $\simeq$ $\gamma$ =  0.026 $\pm$ 0.003.

\subsection{Extended Radio Emission Near \hs}
VLA observations of the gravitationally lensed quasars \hs, RX~J0911$+$0511, HE~0435$-$1223 and SDSS~J0924$+$0219 have revealed radio emission in these radio-quiet objects (Jackson et al. 2015).
Of particular interest is the VLA image of \hs\ that shows a faint arc around images A and B likely formed by extended emission near the quasar and lensed by the intervening galaxy. Jackson et al. model this arc and infer an extended radio source size of about 10~mas that corresponds to a physical size of about 90~pc.  
This is close to the predicted size of the inner portion of the narrow line region (NLR) for an AGN with the luminosity of \hs. 

To test whether the ultrafast outflow is energetic enough to produce the observed radio emission of \hs\ we compare the unlensed radio luminosity of \hs\ at 1.4~GHz with the kinetic luminosity of the wind.  We find the luminosity at 1.4 GHz to be {$\nu$}L$_\nu$~$\sim$~1~$\times$~10$^{39}$~erg~s$^{-1}$ and 
the kinetic luminosity of the outflow to lie in the range 5--30 $\times$ 10$^{46}$ erg~s$^{-1}$. 
If the radio synchrotron emission of \hs\ is produced during the acceleration of particles of the AGN outflow as it collides with the surrounding medium we find that the efficiency 
of conversion is significantly less than that of supernova-driven winds. 

Zakamska et al. 2015 find from the analysis of a sample of obscured radio-quiet quasars that their radio luminosities are correlated with the velocity of outflowing narrow-line gas. They suggest that the weak radio emission of radio-quiet quasars may be produced by particles accelerated in the shocks within the quasar-driven outflows.
Comparing with Figure 10 of Zakamska et al., that plots the radio luminosities of a sample of radio-quiet AGN versus the velocity width of the [O III] emission, we find that the radio luminosity of \hs\ falls on the borderline between the radio luminosities of Seyfert galaxies and radio-quiet quasars.

\section{DISCUSSION}

The 100~ks {\it Chandra} and 46~ks {\it XMM-Newton} observations of the NAL quasar \hs\  confirm the presence of a highly ionized and relativistic outflow in this quasar and provide tight constraints on the properties of the wind. These improved constraints allow more accurate estimates of the mass outflow rate and the rate of kinetic energy injection than the ones provided with the previous 5~ks {\it Chandra} observation.

Our analysis of the P-Cygni profile detected in the {\it Chandra} spectrum of image (A+B) constrains the covering factor 
of the wind to be $f_{\rm c}$= 0.6$^{+0.2}_{-0.3}$. As we discuss later on in this section this value is likely a lower limit and it is subject to a few caveats.
 This large covering factor is consistent with studies of NAL quasars 
(e.g., Misawa et al. 2007; Simon et al. 2012; Culliton et al. 2012) that indicate the true fraction of quasars with 
a quasar-driven outflow responsible for producing NALs to be $\simgt$~40\%.  
We are assuming that the UV and X-ray observations are sampling different parts of the outflow with the X-ray absorbers located 
closer to the continuum source, thus having a higher ionization level than the UV absorbers that are likely located further out.
The detection of a \SiXIV\ resonance absorption line in \hs\ blueshifted by the same amount as that of the \FeXXV\ resonance absorption line
support the presence of a relativistic outflow (see Figures 6 and 8).

In Table 6 we list the 0.2--2~keV and 2--10~keV fluxes and luminosities from the combined images of \hs\ since the images are not resolved with {\sl XMM-Newton}. We find significant flux variability. Specifically, the 0.2$-$2~keV flux increased from 4.0$_{-0.4}^{+0.3}$ $\times$ 10$^{-13}$~erg~s$^{-1}$~cm$^{-2}$
to 6.4$_{-0.3}^{+0.3}$ $\times$ 10$^{-13}$~erg~s$^{-1}$~cm$^{-2}$ between the December 2013 and October 2014 observations. The increase in the 0.2--2~keV flux is accompanied by a significant decrease of the intrinsic neutral column density (see Table 3 and Figure 5).  Despite this significant change of the neutral intrinsic column density the velocity of the outflow does not appear to have changed.We conclude that the soft X-ray flux variability is possibly caused by a change in the intrinsic neutral column density.

For estimating the mass outflow rate of the wind we assume a spherically symmetric wind with a covering factor of $f_{\rm c, i}$ (e.g., Lamers \& Cassinelli 1999, implemented in XSPEC as the \verb+windabs+ model, whose assumptions we describe below).  We approximate $N_{\rm H} \sim n(r){\Delta}r$, where $n(r)$ is the number density of the gas.
We use the following expressions to estimate the mass-outflow rate and the efficiency of outflow:

\begin{equation}
 \dot{M_{\rm i}} = 4{\pi}r_{\rm i}(r_{\rm i}/{\Delta}{r_{\rm i}})N_{\rm H,i}m_{\rm p}v_{\rm wind,i}f_{\rm c,i}
\end{equation}

\begin{equation}
\epsilon_{\rm k,i} = {{1}\over{2}}{\dot{M}_{\rm i}{v^{2}_{\rm wind,i}}\over{L_{\rm Bol}}} 
\end{equation}

\noindent
where ${\Delta}{r_{\rm i}}$ is the thickness of the absorber at radius $r_{\rm i}$, 
$N_{\rm H,i}$ is the hydrogen column density,
$v_{\rm wind,i}$ is the outflow velocity of the X-ray absorber, $f_{\rm c,i}$
is the global covering factor of the absorber, $i$ indicates the absorbing component and $L_{\rm Bol}$ is the bolometric photon luminosity of the quasar.

Estimates of the properties of absorbing outflows in \hs\ for components detected at $>$ 99\% confidence are presented in Table 7.
We base our outflow estimates for the {\sl Chandra} and {\it XMM-Newton} epochs on the best-fit results of model 5 (Table 4)
and model 7 (Table 5), respectively. 

The outflow velocities in the {\sl Chandra} and {\sl XMM-Newton} observations of \hs\ are similar if one considers the estimated error bars.  
Specifically, the two outflow components detected in the {\sl Chandra} observation have 
$v_{\rm abs,1}$ = 0.10$^{+0.01}_{-0.01}$$c$ and $v_{\rm abs,2}$=0.50$^{+0.14}_{-0.08}$$c$ (component 2 is only marginally detected in model 2)  and the two outflow components detected in the {\sl XMM-Newton} observation have
$v_{\rm abs,1}$ = 0.12$^{+0.02}_{-0.01}$$c$ and $v_{\rm abs,2}$=0.41$^{+0.07}_{-0.04}$$c$. The significance of the detections of the absorption lines in the {\sl Chandra} and {\sl XMM-Newton} observations are shown in the confidence contour plots of figures 6 and 10, respectively.

To estimate the bolometric luminosity of \hs\ we correct for the lensing magnification and apply a bolometric correction to the 2--10~keV luminosities based on the empirical relations presented in Lusso et al. 2012. The lensing magnification was determined by modeling the gravitational lens system \hs\ using the gravitational lens adaptive-mesh fitting code glafic version 1.1.6 (Oguri 2010). 
The lens was modelled as a singular isothermal ellipsoid plus an external shear from the nearby galaxy group.
We incorporate the more accurate values of  $\kappa$ and $\gamma$ obtained from the 100~ks observation in our lens model to better constrain the image magnifications, the unlensed luminosity of HS 0810+2554 and the image time-delays. We find that the magnifications of the images are $\mu_{\rm A}$$\sim$42, $\mu_{\rm B}$$\sim$43, $\mu_{\rm C}$$\sim$11, and $\mu_{\rm D}$$\sim$7.
The time-delays between images  are found to be $td_{\rm {BA}}$=86~s, $td_{\rm BC}$=6,740~s, 
$td_{\rm AC}$=6,653~s, and $td_{\rm DC}$=25,833~s.
The mean Eddington ratio for an AGN with the black hole mass and redshift of  \hs\ is $\lambda_{\rm Edd}$~$\sim$~0.1.
Lusso et al. 2012 indicate that the 2--10~keV bolometric correction for type I AGN with an Eddington ratio of $\lambda_{\rm Edd}$~$\sim$~0.1
is $\kappa_{\rm 2-10keV}$~$\sim$~30. The bolometric luminosity of \hs\  during the 100~ks {\it Chandra} observation is $L_{\rm Bol}$=1.1(85/$\mu_{\rm A+B}$)($\kappa_{\rm 2-10keV}$/30)~$\times$~10$^{45}$erg~s$^{-1}$, where the lensing magnification of image (A+B) is $\mu_{\rm AB}$ = 85 and the bolometric correction is $\kappa_{\rm 2-10keV}$=30 for an unlensed 2-10~keV luminosity of 3.73 $\times$~10$^{43}$erg~s$^{-1}$.
The bolometric luminosity of \hs\  during the 46~ks {\it XMM-Newton} observation is $L_{\rm Bol}$~=~1.4(103/$\mu_{\rm ABCD}$)($\kappa_{\rm 2-10keV}$/30)~$\times$~10$^{45}$erg~s$^{-1}$.

We note that the bolometric luminosity of \hs\ is not well constrained because of the large uncertainty in the 2--10~keV bolometric correction.
An independent estimate of $L_{\rm Bol}$ is provided from the monochromatic luminosities of \hs\ at 1450$\rm \AA$ and 5100$\rm \AA$ 
based on the empirical equations of Runnoe et al. (2012).
Assef et al.  (2011) list the continuum monochromatic luminosities  $\log({\lambda}L/\rm erg~s^{-1})$ of \hs\ at 1450$\rm \AA$ and 5100$\rm \AA$ to be 
44.44 and 44.84, respectively.  
Using the recommended bolometric corrections of Runnoe et al. we estimate the bolometric luminosities based on the  1450$\rm \AA$ and 5100$\rm \AA$  continuum monochromatic luminosities of \hs\  to be 1.1 $\times$~10$^{45}$erg~s$^{-1}$ and 3.7 $\times$~10$^{45}$erg~s$^{-1}$, respectively.
For the purpose of estimating the efficiency of the outflow we adopt a mean value for the bolometric luminosity of $L_{\rm Bol}$~=~1.8 $\pm$ 1.3 $\times$~10$^{45}$erg~s$^{-1}$,
based on the X-ray, UV, and optical bolometric corrections. We assumed a global covering factor of the outflowing absorber of $f_{\rm c}$=0.6$_{-0.3}^{+0.2}$ based on our fits to the P-Cygni profile detected in the {\sl Chandra} spectrum of \hs.

For estimating the mass outflow rate and outflow efficiency, we assumed a fraction  $r/{\Delta}{r}$ ranging from 1 to 10 based on 
theoretical models of quasar outflows (e.g., Proga et al. 2000). Assuming that the maximum outflow 
velocity is produced by gas that has reached its terminal velocity one obtains the approximation
$R_{\rm launch}$ $\sim$ a few times $R_{s \rm}(c/v_{\rm wind} )^{2}$ , where $v_{wind}$ is the observed outflow velocity, $R_{\rm launch}$ is the launch radius of the outflow,  and $R_{\rm s}$ = $2GM/c^2$. Based on our estimated maximum outflow velocity ($v~\sim~0.4c$), we expect 
$r$ to be similar to  $R_{\rm launch}$ and range between 3~$R_{\rm s}$ and 20~$R_{\rm s}$.
Our assumed range of $r$ is conservative, ranging from the innermost stable circular orbit (ISCO) radius to 20~$R_{\rm s}$.
Launching a wind at significantly smaller radii than the ISCO radius is unlikely
since general relativistic (GR) effects on an absorber launched within the ISCO radius would result in significant
redshifts of the absorption lines and launching radii greater than 20~$R_{\rm s}$ will result in even
larger values of the mass-outflow rate assuming the mass-outflow rate is given by the expression presented in Equation 1.
We note, however, that in Equation 1 we approximate $N_{\rm H}$ $\sim$ $n(r)$${\Delta}$r, assume a continuous wind smoothly distributed over the disk, and
do not consider velocity gradients of the wind or include special relativistic corrections (see Equation 4 of Saez \& Chartas, 2011).

We used a Monte Carlo approach to estimate the errors of $\dot{M}_{\rm i}$ and $\epsilon_{\rm k}$.
The values of $v_{\rm wind}$ and $N_{\rm Habs}$ were assumed to have normal distributions within their error limits.
The values of $f_{\rm c}$, $r/{\Delta}{r}$, and $r$ were assumed to have uniform distributions within their error limits.
By multiplying these distributions and with the appropriate constants from equations 2 and 3 we obtainedÊ
the distributions of $\dot{M}_{\rm i}$ and $\epsilon_{\rm k}$.
We finally determined the mean values of the distributions of 
$\dot{M}_{\rm i}$ and $\epsilon_{\rm k}$ and estimated the 68\% confidence ranges.

In Table 7 we list the total hydrogen column densities $N_{\rm H}$ of the X-ray absorption lines,
the outflow velocity of each absorption component, the mass-outflow  rates and the efficiency of the outflows.
The large fraction of kinetic to electromagnetic luminosity implies that radiation driving alone cannot explain the acceleration of this highly-ionized absorber. 
Magnetic driving likely provides a significant contribution to the acceleration of the X-ray wind (e.g., Kazanas et al. 2012;  Fukumura et al. 2014).
The mass outflow rate of \hs\ is found to lie in the range 1.5-3.4 $M_{\odot}$~yr$^{-1}$ which is comparable to the 
accretion rate of \hs\  which we estimate
to be 1.8 $\times$ 10$^{-3}$($L_{44}$/$\eta$)$M_{\odot}$  yr$^{-1}$ $\sim$~1~$M_{\odot}$~yr$^{-1}$,
where we assumed a typical accretion efficiency of $\eta$ = 0.1.
For calculating the mass-outflow rate we have assumed a continuous biconical outflow with a covering factor. Another possibility is that the wind is not continuous but made up of clouds. We note that  Equation 1 is approximately valid in the case of clouds too, as long as the clouds are distributed throughout the shell of thickness ${\Delta}$r. Future observations with the X-ray satellite {\sl Hitomi} with its high energy resolution may be able to show whether the blueshifted broad X-ray absorption features are caused by a velocity gradient in a continuous wind. The confirmation of smooth broad absorption lines would suggest that the outflow is  made of a continuous stream of ionized material instead of an outflow of dense clouds.

The {\sl XMM-Newton} spectrum of \hs\ shows evidence of absorption and reflection (see Figure 4). Model 6 assumes a distant reflector and two outflowing absorbers, 
whereas, model 7 includes a nearby reflector and two outflowing absorbers. 
In both cases the reflection component lies below the direct component except for the rest-frame energy range of 15$-$50 keV. 
If we define the reflection fraction, $R$, as the ratio of the flux reflected by the disk to the
flux of the direct power-law component in the rest-frame energy band of 20$-$40~keV, we find that $R = 1.4 \pm 0.6$ (for model 7 of Table 5).
If light-bending of the direct component is not important then the direct component should in general  be larger than the reflected one.
However, several recent observations and simulations indicate that the reflection component can dominate in some AGN and the relative strength of the reflected versus direct X-ray emission increases with the spin of the black hole.
In particular, recent studies of AGN have attempted to disentangle the direct from the reflected X-ray emission (e.g., Fabian 2014; Zoghbi et al. 2015; Gallo et al. 2015; Keck et al. 2015).  To complicate the analysis further the reflected X-ray emission of AGN is thought to be comprised of two main components; (a) backscattering from distant cold material in a dusty molecular torus or in the  broad line region gas and (b) backscattering from ionized disk material near the black hole. The reflected emission from the inner disk is expected to be blurred due to relativistic effects. The  reflection fraction in some cases is found to be larger than 1 (e.g, Zoghbi et al. 2015; Gallo et al. 2015; Keck et al. 2015).  Reflection dominated X-ray spectra can be caused by light-bending of a portion of the direct emission component form the hot corona into the black hole resulting in an increase in the reflection fraction. Dauser et al. 2014 show through simulations that reflection fractions in excess of 2 are only possible for rapidly spinning black holes. 

There are additional effects that may distort the X-ray emission from the corona such as scattering off an outflowing wind and absorption by material near the black hole (e.g., Sim et al. 2012; Miller et al. 2009).  The spectral contribution from scattering off outflowing X-ray absorbers has been simulated in Sim et al. 2012. In their simulations (see figures 6 and 7 of Sim et al. 2012) 
the scattering off the wind affects mostly the spectral region near the Fe K${\alpha}$ line  (6.4$-$7~keV) but does not produce an excess at higher rest-frame energies (10$-$100~keV) similar to that produced by backscattering from the optically thick accretion disk. The Sim et al. simulations also indicate the contribution from scattering to be inclination-angle dependent.  Specifically, the spectral distortion caused by scattering off the wind near the iron line region decreases with decreasing inclination angle. We therefore conclude that the hard excess observed above rest-frame energies of $\sim$12~keV in the {\sl XMM-Newton} spectrum of \hs\ is not the result of scattering off the outflowing X-ray absorber but likely the result of backscattering from the accretion disk. 

There are several assumptions that are made in the \verb+windabs+ model that we used to fit the P-Cygni profile of the spectrum of image A+B.
\verb+windabs+ assumes a spherical and isotropic outflow ionized by a compact central continuum source  and with a  covering factor $f_{\rm c}$.
In the case of \hs\ the central ionizing source of the outflowing X-ray absorbing material is the hot corona. Microlensing studies of radio-quiet quasars with black-hole masses ranging between $\sim$10$^{7}$ M$_{\odot}$ and $\sim$10$^{9}$ M$_{\odot}$ indicate that their hot coronae are compact with sizes close to 6 r$_{\rm g}$ (Chartas et al. 2016).
We therefore consider as a reasonable assumption that the ultrafast wind of \hs\  is being illuminated by a compact central continuum source.
The kinetic and geometric structure of ultrafast winds in radio-quiet quasars in not well known. 
A departure of the outflow from isotropy will in general distort the shape of the reflected component (emission line) of the P-Cygni profile.
\verb+windabs+  does not account for the presence of the accretion disk which may shield a significant portion of the reflected resonant emission from the wind.
Dorodnitsyn 2009 has simulated the effect of shielding of an outflow by the AGN accretion disk and finds that for small inclination angles shielding results in significant attenuation
 of the red wing of the emission line of the P-Cygni profile (see Figure 12 of Dorodnitsyn 2009). This suggests that the value of the covering factor obtained in our analysis using \verb+windabs+  is likely an underestimate and thus provides a lower limit of the true covering factor. We reproduced this effect by simulating a spectrum with a P-Cygni profile with an emission line of twice the observed strength (unblocked case) and fitted the spectrum using the \verb+windabs+ model. The covering factor in this case increased by a factor of 2.
We conclude that the covering factor of the wind in \hs\ is likely large with  $f_{\rm c}$ $\simgt$ 0.6. Future observations with the high-energy resolution calorimeter onboard the X-ray satellite {\sl Hitomi}
should be able to resolve the structure of the P-Cygni profile and provide constraints on disk shielding and the presence of general relativistic effects on the absorption profile.

The fits to the {\sl XMM-Newton} spectrum of \hs\ using models 6 and 7 that include a P-Cygni profile do not place any useful constraints on the opening angle of the wind. Scattering from the outflowing X-ray absorber is expected to produce an emission line near the energy of the Fe~K$\alpha$ line but it is not detected at a high significance level in the {\sl XMM-Newton} spectrum as it is in the {\sl Chandra} spectrum of image (A+B). This could in part be caused by the fact that the {\sl XMM-Newton}  spectrum combines all images and the P-Cygni profile may be smeared if the spectra of images A+B and C differed during the observation. Another possible reason is variability of the covering fraction. There are hints of such variability of the covering fraction in the ultrafast outflow of PDS~456 (see Figure 1 and Table S2 in Nardini et al. 2015).

We have not investigated more complex spectral models that include multiple reflectors and scatterers due to the relatively low S/N of the available X-ray spectra of \hs\ and the lack of spectral coverage in a higher energy range where the Compton reflection component is expected to peak. A scheduled joint observation of \hs\ with {\it Chandra} and {\it NuSTAR} will provide improved constraints of the kinematic and ionization properties of the ultrafast outflow and reflector/s of HS~0810+2554 by fitting outflow/reflection models  to the 0.3--80 keV
spectrum of \hs.

\section{SUMMARY}

The 100~ks {\sl Chandra} and 46~ks {\sl XMM-Newton} observations of the gravitationally lensed $z$ = 1.51 narrow absorption line quasar  \hs\
indicate (at $>$ 99\% confidence) the presence of a highly ionized and relativistic outflow in this highly magnified object. These observations
confirm our marginal detection of an ultrafast wind in this object based on an earlier 5~ks {\it Chandra} observation. 

We summarize the main results from the latest X-ray observations of \hs\ as follows:

1. A highly ionized and relativistic wind is present in both the 100~ks {\it Chandra} and 46~ks {\it XMM-Newton} observations of \hs. 
We find the ionization parameter of the X-ray outflowing absorber to lie in the range $\log$($\xi$/erg~cm~s$^{-1}$) =  3.2$-$3.3,
the hydrogen column density to lie within the range  $N_{\rm Habs}$ = 2.9--3.4 $\times$ 10$^{23}$~~cm$^{-2}$,
and the outflow velocity components to lie within the range  $v_{\rm abs}$ = 0.1$-$0.41 $c$.

2. Our interpretation of the relativistic nature of the outflow is consistent with the detection of a \SiXIV\ absorption line 
blueshifted by the same amount as the highly ionized Fe line in the 100~ks {\it Chandra} spectrum of image A+B.
The presence of an outflow is also supported by the detection of a P-Cygni profile in the 100~ks {\it Chandra} spectrum of image A+B.
We constrain the covering factor $f_{\rm c}$ of the relativistic wind by fitting the P-Cygni profile. We find $f_{\rm c}$ = 0.6$^{+0.2}_{-0.3}$ indicating that the wind is wide-angled and
therefore important in the feedback process by providing efficient coupling of the outflow to the surround intergalactic medium.
The value of the covering factor obtained in our analysis using \verb+windabs+ is likely an underestimate because of shielding of the outflow by the accretion disk and thus provides a lower limit of the true covering factor. We conclude that the covering factor of the wind in HS~0810+2554 is likely large with $f_{\rm c}$ $\simgt$~0.6, assuming disk shielding.

3. A hard X-ray excess component is detected in the 46~ks {\it XMM-Newton} spectrum of \hs. 
A possible explanation is reflection from nearby and/or distant regions of the accretion disk. 
By modelling the hard X-ray excess as blurred relativistic reflection from the accretion disk we constrain the
inclination angle of the disk to be $\simlt$ 45$^{\circ}$ (90\% confidence). We note that this result is model dependent (see Table 5) and that due to the relatively low S/N
of the available spectra of \hs\ several parameters of model 7 (see Table 5) were fixed 
during the spectral fit. A scheduled joint {\sl Chandra} and {\sl NuSTAR} observation of \hs\ will provide tighter constraints on the properties of the reflector.
Such a small inclination angle, if confirmed with the follow-up observations, is consistent with models that posit NAL quasars as objects observed at relative low inclination angles.

4. The mass-outflow rate of \hs\ is found to lie in the range of $\dot{M}$~=~1.5$-$3.4~M$_{\odot}~yr^{-1}$ and is comparable to the accretion rate 
1 M$_{\odot}~yr^{-1}$ (assuming an accretion efficiency of 0.1 and an Eddington ratio of 0.1).
The large fraction of kinetic to electromagnetic luminosity of ($\epsilon_{\rm k}$~=~9$_{-6}^{+8}$) implies that radiation driving alone cannot explain the acceleration of this highly-ionized absorber. Magnetic driving likely provides a significant contribution to the acceleration of the X-ray wind.

5.  The deeper X-ray observations  provide tighter constraints on the properties of a nearby group of galaxies that contributes to the gravitational lensing of \hs.
In particular, joint fits to the {\sl Chandra} and {\it XMM-Newton}  spectra of the group indicate 
a plasma temperature of $T_{\rm e}$ $\sim$ 0.8 keV,  an abundance of A $\sim$ 0.53 solar and a redshift of $z$~$\sim$~0.08.
We used these new constraints to improve our lens model of \hs. We find that the total magnification of the background quasar is $\mu_{ABCD}$ $\sim$ 103.

6. The {\it Chandra}  spectra of images C and A+B differ significantly. 
We propose as a possible explanation for the difference between these spectra a microlensing event occurring in image C. This interpretation is consistent with the anomalous X-ray flux ratio of image C (see Table 4).
Microlensing may lead to energy shifts of the iron emission lines due to general and relativistic effects as found in the lensed quasar \rxj (Chartas et al. 2016).

\acknowledgments
We acknowledge financial support from NASA via the Smithsonian Institution grants SAO GO1-12146B.
ME acknowledge financial support from NSF grant AST-0807993.

\clearpage
\begin{table}
\caption{Log of Observations of Quasar \hs\ } 
\scriptsize
\begin{center}
\begin{tabular}{lccccccccccc}
                           &                        &   &   & & & & & &\\ \hline\hline
                           &                     &                      & Exposure            &                                                               &                                                            &                                                             &  & \\
 Observation      & Observatory &  Observation &   Time$\tablenotemark{a}$  &  $N_{\rm A}$$\tablenotemark{b}$  & $N_{\rm B}$$\tablenotemark{b}$ & $N_{\rm C}$$\tablenotemark{b}$ & $N_{\rm D}$$\tablenotemark{b}$ & $N_{\rm ABCD}$ $\tablenotemark{b}$ \\
     Date              &                      &        ID          &   (s)                                      & counts                                            & counts                                               & counts                                                & counts & counts   \\
\hline
&   &  &  &   & & & &\\
2002 January  30  & {\sl Chandra}            & 3023             &  4,894                  &  290$^{+33}_{-27}$                             & 239$^{+41}_{-43}$                      & 145$^{+15}_{-18}$            & 34$^{+9}_{-8}$         & 708 $\pm$ 27 \\ 
2013 December 16 & {\sl Chandra}          & 16110            &  97,730                &  2,179$^{+92}_{-127}$                        & 1,463$^{+91}_{-82}$                    & 1,723$^{+62}_{-60}$            & 411$^{+26}_{-25}$ &5,776 $\pm$ 76 & \\ 
2014 October 4       & {\sl XMM-Newton}  &0728990101   &  46,400                & ---                                                       & ---                                                 & ---                                        & ---                        & 11,810 $\pm$ 115\\
&   &  &  &   & & & &\\
\hline 
\end{tabular}
\end{center}
${}^{a}${Exposure time is the effective exposure time remaining after the application of good time-interval (GTI) tables and the removal of portions of the
observation that were severely contaminated by background flaring.}\\
${}^{b}${Background-subtracted source counts including events with energies within the 0.2$-$10 keV band. The source counts and effective exposure
times for the {\sl XMM-Newton} observation refer to those obtained with the EPIC PN instrument.}\\
\end{table}

\clearpage
\begin{table}
\caption{Multi-Wavelength Flux Ratios of \hs\ Components}
\scriptsize
\begin{center}
\begin{tabular}{ccccc}
 & & & &\\ \hline\hline
\multicolumn{1}{c} {Waveband} &
\multicolumn{1}{c} {A/Total} &
\multicolumn{1}{c} {B/Total} &
\multicolumn{1}{c} {C/Total} &
\multicolumn{1}{c} {D/Total}\\ \hline
 & & & &\\
H band           & 0.52 $\pm$ 0.01 & 0.30 $\pm$ 0.01 & 0.14 $\pm$ 0.01 &  0.041 $\pm$ 0.002\\
I band           & 0.58 $\pm$ 0.05 & 0.26 $\pm$ 0.09 & 0.13 $\pm$ 0.03  &  0.04 $\pm$ 0.01 \\
V band           & 0.52 $\pm$ 0.03 & 0.25 $\pm$ 0.04 & 0.18 $\pm$ 0.03 & 0.05 $\pm$ 0.01 \\
0.2$-$10~keV{}$^{a}$           & 0.41 $\pm$ 0.04 & 0.34 $\pm$ 0.06 & 0.21 $\pm$ 0.02 & 0.05 $\pm$ 0.02  \\ 
0.2$-$10~keV{}$^{b}$           & 0.38 $\pm$ 0.02 & 0.25 $\pm$ 0.02 & 0.30 $\pm$ 0.01 & 0.07 $\pm$ 0.01 \\ 
 & & & &\\
\hline 
\end{tabular}
\end{center}
\noindent
\tablecomments{The H, I, and V band data are taken from 
the CfA-Arizona Space Telescope LEns Survey (CASTLES) of gravitational lenses
website {\it http://cfa-www.harvard.edu/glensdata/}.
Error bars for the X-ray data are at the 68\% confidence level.\\
{}$^{a}$ X-ray flux ratios for the 2002 observation. \\
{}$^{b}$ X-ray flux ratios for the 2013 observation.}
\end{table}

\clearpage
\begin{table}
\caption{Results from Fits  to \chandra\ and \xmm\ Spectra of \hs}
\scriptsize
\begin{center}
\begin{tabular}{ccccc}
 & & & &   \\ \hline\hline
\multicolumn{1}{c} {Model$^{a}$} &
\multicolumn{1}{c} {Parameter$^{b}$} &
\multicolumn{1}{c} {Fitted Values${}^{c}$} & 
\multicolumn{1}{c} {Fitted Values${}^{c}$}  &
\multicolumn{1}{c} {Fitted Values${}^{c}$}  \\
        &              &  Image (A+B)     & Image C          &    Image (A+B+C+D)       \\
        &              & \chandra\      &   \chandra\       &    \xmm\            \\ 
         \hline
           &  &                     &           \\
  ${1}$ &$\Gamma$        &  1.96$_{-0.08}^{+0.09}$                                                   &    1.91$_{-0.10}^{+0.10}$  &  2.08$_{-0.03}^{+0.04}$   \\
    &   $N_{\rm H}$           &  0.3$_{-0.16}^{+0.20}$$\times$10$^{22}$~cm$^{-2}$   &  $<$ 0.2$\times$10$^{22}$~cm$^{-2}$                        & $<$ 0.4$\times$10$^{22}$~cm$^{-2}$   \\
 &$\chi^2/{\nu}$               & 97.9/105                                                                         &  90.0/70                                                                        & 259/206  \\
&$P(\chi^2/{\nu})$$^{d}$ & 6.8~$\times$~10$^{-1}$                                                    & 5.4~$\times$~10$^{-2}$                                              &  7.2~$\times$~10$^{-3}$  \\  
    &  &                     &           \\
${2}$ &  $\Gamma$            & 1.83$_{-0.07}^{+0.06}$                                                   &    1.97$_{-0.15}^{+0.09}$      & 2.25$_{-0.03}^{+0.04}$ \\
&   $N_{\rm H}$                  & 0.15 $_{-0.05}^{+0.13}$$\times$10$^{22}$~cm$^{-2}$  &  0.21 $_{-0.18}^{+0.18}$$\times$10$^{22}$~cm$^{-2}$  & $<$ 0.04$\times$10$^{22}$~cm$^{-2}$ \\
  &  E$_{\rm 1}$           & 6.8$_{-0.2}^{+0.2}$~keV                                                    &    1.76$_{-0.28}^{+0.05}$~keV        &  7.89$_{-0.31}^{+0.22}$~keV \\
  &  $\sigma_{\rm 1}$  & 0.3$_{-0.1}^{+0.2}$~keV                                                    &    $ < $ 0.24~keV                             & 0.26$_{-0.22}^{+0.30}$~keV  \\
  &  EW$_{\rm 1}$       & $+$465$_{-286}^{+334}$~eV                                           &  $+$114$_{-114}^{+115}$~eV       & $-$272$_{-95}^{+88}$~eV \\
  &  E$_{\rm 2}$            & 2.19$_{-0.1}^{+0.1}$~keV                                                 &  5.87$_{-0.28}^{+0.31}$~keV      &   10.9$_{-0.6}^{+0.7}$~keV\\
  &  $\sigma_{\rm 2}$    & $<$ 0.3~keV                                                                    &  0.24$_{-0.24}^{+0.27}$~keV       & 0.88 $_{-0.39}^{+1.02}$~keV \\
  &  EW$_{\rm 2}$         &$-$48$_{-25}^{+28}$~eV                                                  &    $+$327$_{-327}^{+312}$~eV     &  $-$702$_{-229}^{+230}$~eV    \\
   &  E$_{\rm 3}$           & 7.79$_{-0.51}^{+0.53}$~keV                                            &    8.81$_{-0.15}^{+0.28}$~keV     & ---      \\
  &  $\sigma_{\rm 3}$   &  0.91$_{-0.49}^{+0.59}$~keV                                                 &  $ < $ 0.5 ~keV                      & ---    \\
  &  EW$_{\rm 3}$        & $-$721$_{-225}^{+195}$~eV                      &    $+$457$_{-200}^{+212}$~eV       &---     \\
  &  E$_{\rm 4}$            & 13.1$_{-1.51}^{+1.57}$~keV                    &     ---  & --- \\
  &  $\sigma_{\rm 4}$   & 1.92$_{-1.53}^{+5.1}$~keV                      &  ---  & --- \\
  &  EW$_{\rm 4}$        & $-$1570$_{-680}^{+650}$~eV                   &     ---    & ---    \\
&     A$_{\rm Fe}$         &--- &  ---      & 1 A$_{\odot}$  (fixed)                \\
&     E$_{\rm cutoff}$     &--- &  ---              & 1000~keV  (fixed)                \\
&      Inclination              &--- &   ---    & $ < $ 50 degrees (68 $\%$  confidence)            \\
  & $\chi^2/{\nu}$                & 68.1/93                                                                   & 73.1/61                                                   & 214.5/197 \\
 &$P(\chi^2/{\nu})$$^{d}$   & 9.8~$\times$~10$^{-1}$                                      &  1.4~$\times$~10$^{-1}$                               & 1.9~$\times$~10$^{-1}$   \\
&  &                      &           &\\
 \hline 
\end{tabular}
\end{center}
\noindent
${}^{a}$Model 1 consists of a power law and neutral absorption at the source. 
Model 2  consists of a power law, neutral absorption at the source, and Gaussian absorption and/or emission lines at the source.
Model 2 fit to the {\sl XMM-Newton} spectrum also includes neutral Compton reflection with self-consistent Fe and Ni lines (PEXMON). 
All model fits include the Galactic absorption toward the source (Dickey \& Lockman 1990).\\
${}^{b}$All absorption-line parameters are calculated for the rest frame.\\
${}^{c}$Spectral fits were performed using the $\chi^{2}$ statistic and all errors are for 90\% confidence unless mentioned otherwise. \\
${}^{d}$$P(\chi^2/{\nu})$ is the probability of exceeding $\chi^{2}$ for ${\nu}$ degrees of freedom
if the model is correct.\\
\end{table}

\clearpage
\setcounter{table}{2}
\begin{table}
\caption{\text{(Continued)} }
\scriptsize
\begin{center}
\begin{tabular}{ccccc}
 & & & &   \\ \hline\hline
\multicolumn{1}{c} {Model$^{a}$} &
\multicolumn{1}{c} {Parameter$^{b}$} &
\multicolumn{1}{c} {Fitted Values${}^{c}$} & 
\multicolumn{1}{c} {Fitted Values${}^{c}$}  &
\multicolumn{1}{c} {Fitted Values${}^{c}$}  \\
        &              &  Image (A+B)     & Image C          &    Image (A+B+C+D)       \\
        &              & \chandra\      &   \chandra\       &    \xmm\            \\  \hline
    &           &                                    &   &\\
${3}$  &$\Gamma$            & 2.03$_{-0.08}^{+0.09}$                                                      &  1.94$_{-0.05}^{+0.15}$                                     & 2.20$_{-0.04}^{+0.04}$ \\
&   $N_{\rm H}$                  & 0.39 $_{-0.13}^{+0.15}$$\times$10$^{22}$~cm$^{-2}$ &  $<$ 0.37 $\times$10$^{22}$~cm$^{-2}$            & $<$ 0.03$\times$10$^{22}$~cm$^{-2}$  \\
 &  E$_{\rm 1}$           & 6.96$_{-0.24}^{+0.25}$~keV              &    1.78$_{-0.16}^{+0.08}$~keV        &  --- \\
  &  $\sigma_{\rm 1}$  & 0.38$_{-0.26}^{+0.65}$~keV              &    $ < $ 0.20~keV                             & --- \\
  &  EW$_{\rm 1}$       & $+$582$_{-376}^{+327}$~eV             &  $+$105$_{-90}^{+96}$~eV         & ---  \\
 &  E$_{\rm 2}$           & ---                                                        &    5.88$_{-0.29}^{+0.55}$~keV        &  --- \\
  &  $\sigma_{\rm 2}$  & ---                                                        &    0.23$_{-0.23}^{+0.45}$~keV      & ---  \\
  &  EW$_{\rm 2}$       & ---                                                        &  $+$282$_{-237}^{+280}$~eV       & ---  \\
 &  E$_{\rm 3}$           & ---                                                          &    8.81$_{-0.28}^{+0.05}$~keV      & ---  \\
  &  $\sigma_{\rm 3}$  & ---                                                          &    $ < $ 0.46~keV                          & ---  \\
  &  EW$_{\rm 3}$       & ---                                                         &  $+$417$_{-172}^{+177}$~eV       & --- \\
  &   $N_{\rm Habs1}$        & 3$_{-1.5}^{+1.8}$ $\times$ 10$^{23}$~cm$^{-2}$     &   0.9$_{-0.9}^{+0.4}$ $\times$ 10$^{23}$~cm$^{-2}$   & 5$_{-3}^{+4}$ $\times$ 10$^{23}$~cm$^{-2}$\\
  &    $\log\xi_{\rm abs1}$          & 3.25$_{-0.2}^{+0.35}$~erg~cm~s$^{-1}$               &  3.1$_{-0.3}^{+0.2}$~erg~cm~s$^{-1}$(68\%)         &3.5$_{-0.3}^{+0.3}$~erg~cm~s$^{-1}$\\
& $v_{\rm turb,abs1}$         &   22,000 km~s$^{-1}$                                                &  5,500 km~s$^{-1}$                                                                    &18,500 km~s$^{-1}$ \\
   &   $z_{\rm abs1}$                  &  1.26$_{-0.06}^{+0.07}$                                          &  1.49$_{-0.21}^{+0.13}$                                       &  1.20$_{-0.07}^{+0.07}$\\
  &   $N_{\rm H,abs2}$        & ---                                                                             &   ---                                                                              & 4.5$_{-2.1}^{+1.9}$ $\times$ 10$^{23}$~cm$^{-2}$ \\
  &    $\log\xi_{\rm abs2}$          & ---                                                                             &  ---                                                                          &  3.3$_{-0.2}^{+0.4}$~erg~cm~s$^{-1}$\\
& $v_{\rm turb,abs2}$         &          ---                                                                     &   ---                                                                                         & 30,000 km~s$^{-1}$ \\
 &   $z_{\rm abs2}$                  &  ---                                                                           &  ---                                                                             & 0.60$_{-0.08}^{+0.07}$\\
&     A$_{\rm Fe}$         &--- &---       & 1 A$_{\odot}$  (fixed)                \\
&     E$_{\rm cutoff}$     &  ---        & ---      & 1000~keV  (fixed)                \\
&      Inclination              &--- & ---     & $ < $ 50 degrees (68 $\%$  confidence)            \\
&   $\chi^2/{\nu}$               & 70.7/99                                                                    &  71.2/58                                                                          & 212.1/197 \\
&  $P(\chi^2/{\nu})$$^{d}$ & 9.9~$\times$~10$^{-1}$                                            &   1.1~$\times$~10$^{-1}$                                           &2.2~$\times$~10$^{-1}$\\ 
&  &                      &           &\\
  \hline
\end{tabular}
\end{center}
\noindent
${}^{a}$
Model 3 consists of a power law, neutral absorption at the source and one or two outflowing ionized absorbers at the source, and a number of emission lines.
Model 3 fit to the {\sl XMM-Newton} spectrum also includes neutral Compton reflection with self-consistent Fe and Ni lines (PEXMON). 
All model fits include the Galactic absorption toward the source (Dickey \& Lockman 1990).\\
${}^{b}$All absorption-line parameters are calculated for the rest frame.\\
${}^{c}$Spectral fits were performed using the $\chi^{2}$ statistic and all errors are for 90\% confidence unless mentioned otherwise. \\
${}^{d}$$P(\chi^2/{\nu})$ is the probability of exceeding $\chi^{2}$ for ${\nu}$ degrees of freedom
if the model is correct.\\
\end{table}

\clearpage
\begin{table}
\caption{Results from Fits  to the  {\it Chandra} Spectrum of Image (A+B) of  \hs\ with Models that Include a P-Cygni Profile}
\scriptsize
\begin{center}
\begin{tabular}{ccc}
 & &  \\ \hline\hline
\multicolumn{1}{c} {Model$^{a}$} &
\multicolumn{1}{c} {Parameter$^{b}$} &
\multicolumn{1}{c} {Fitted Values${}^{c}$}  \\
        & Image (A+B)           \\ \hline
&  &                     \\
 ${4}$  &$\Gamma$            & 1.97$_{-0.11}^{+0.05}$                          \\
&   $N_{\rm H}$                  & 0.31$_{-0.09}^{+0.15}$$\times$10$^{22}$~cm$^{-2}$    \\
&     $f_{\rm c}$                & 0.58$_{-0.3}^{+0.2}$ (68\%)                   \\                        
  &   N$_{\rm H,abs}$        & 2.7$_{-1.5}^{+1.1}$ $\times$ 10$^{23}$~cm$^{-2}$   \\
  &    $\log\xi_{\rm abs}$          & 3.14$_{-0.06}^{+0.20}$~erg~cm~s$^{-1}$               \\
& $v_{\rm turb,abs}$ &  22,000 km~s$^{-1}$                   \\
   &   $z_{\rm abs}$                  &  1.29$_{-0.03}^{+0.02}$                                         \\ 
&   $\chi^2/{\nu}$               & 81.3/101             \\
&  $P(\chi^2/{\nu})$$^{d}$ & 9.25~$\times$~10$^{-1}$             \\ 
&  &                     \\

${5}$  &$\Gamma$            & 1.98$_{-0.13}^{+0.12}$                          \\
&   $N_{\rm H}$                  & 0.28$_{-0.09}^{+0.31}$$\times$10$^{22}$~cm$^{-2}$    \\
&     $f_{\rm c}$                & 0.6$_{-0.3}^{+0.2}$ (68\%)                 \\
&  E$_{\rm }$           & 6.5$_{-0.2}^{+0.3}$~keV   \\
&  $\sigma$$_{\rm }$           & 1$\times$ 10$^{-3}$~keV (fixed)   \\
 &  EW$_{\rm }$       & $+180$$_{-0.6}^{+150}$~eV   \\
  &   N$_{\rm Habs}$        & 2.9$_{-0.4}^{+1.6}$ $\times$ 10$^{23}$~cm$^{-2}$   \\
  &    $\log\xi_{\rm abs}$          & 3.15$_{-0.21}^{+0.13}$~erg~cm~s$^{-1}$               \\
& $v_{\rm turb,abs}$ &  22,000 km~s$^{-1}$                   \\
   &   $z_{\rm abs}$                  &  1.28$_{-0.03}^{+0.02}$                                         \\ 
&   $\chi^2/{\nu}$               & 77.1/99             \\
&  $P(\chi^2/{\nu})$$^{d}$ & 9.5~$\times$~10$^{-1}$             \\ 
&  &                     \\
  \hline 
\end{tabular}
\end{center}
\noindent
${}^{a}$ Model~4  consists of a power law, neutral absorption at the source, an outflowing ionized absorber at the source
and scattering of the outflowing wind.
The model also includes the Galactic absorption toward the source (Dickey \& Lockman 1990).
Model~5 includes the same components as model 4 with the addition of a narrow emission line near the location of the Fe~K$\alpha$ fluorescent line. \\
${}^{b}$All absorption-line parameters are calculated for the rest frame.\\
${}^{c}$Spectral fits were performed using the $\chi^{2}$ statistic and all errors are for 90\% confidence unless mentioned otherwise. \\
${}^{d}$$P(\chi^2/{\nu})$ is the probability of exceeding $\chi^{2}$ for ${\nu}$ degrees of freedom
if the model is correct.\\
\end{table}

\clearpage
\begin{table}
\caption{Results from Fits  to the  \xmm\ Spectrum of \hs\ with a Model that Includes Reflection and a P-Cygni Profile}
\scriptsize
\begin{center}
\begin{tabular}{ccc}
 & &  \\ \hline\hline
\multicolumn{1}{c} {Model$^{a}$} &
\multicolumn{1}{c} {Parameter$^{b}$} &
\multicolumn{1}{c} {Fitted Values${}^{c}$}  \\
        & Image (A+B+C+D)           \\ 
        \hline
&  &                    \\
 ${6}$  &$\Gamma$            & 2.18$_{-0.07}^{+0.13}$                          \\
&   $N_{\rm H}$                  & 0.03 $_{-0.03}^{+0.07}$$\times$10$^{22}$~cm$^{-2}$    \\
&  Direct Normalization  & 9.6 $\times$ 10$^{-4}$ photons~keV$^{-1}$~cm$^{-2}$~s$^{-1}$ \\
&     A$_{\rm Fe}$                & 1 A$_{\odot}$  (fixed)                \\
&     E$_{\rm cutoff}$                    & 1000~keV  (fixed)                \\
&      Inclination                    & $ < $ 50 degrees (68 $\%$  confidence)            \\
&  Reflection Normalization  & 1.6 $\times$ 10$^{-4}$ photons~keV$^{-1}$~cm$^{-2}$~s$^{-1}$ \\
&  $R_{\rm frac}$                      & 1.1$^{+0.4}_{-0.4}$  (68 $\%$  confidence)   \\
 &   N$_{\rm H,abs1}$        & 2.9$_{-2.0}^{+1.5}$ $\times$ 10$^{23}$~cm$^{-2}$   \\
  &    $\log\xi_{\rm abs1}$          & 3.3$_{-0.15}^{+0.4}$~erg~cm~s$^{-1}$               \\
  & $v_{\rm turb,abs1}$ &    18,500 km~s$^{-1}$ (fixed)                   \\
   & $f_{\rm c,abs1}$ &    $>$ 0.4 (90 $\%$  confidence)                  \\
   &   $z_{\rm abs1}$                  &  1.20$_{-0.04}^{+0.04}$                                         \\ 
   &   N$_{\rm H,abs2}$        & 4.5$_{-2.0}^{+1.8}$ $\times$ 10$^{23}$~cm$^{-2}$   \\
  &    $\log\xi_{\rm abs2}$          & 3.3$_{-0.5}^{+0.5}$~erg~cm~s$^{-1}$ \\
  & $v_{\rm turb,abs2}$ &  30,000 km~s$^{-1}$  (fixed)                  \\
 &   $z_{\rm abs2}$                  & 0.61$_{-0.11}^{+0.11}$  \\
&   $\chi^2/{\nu}$               & 214.4/197             \\
&  $P(\chi^2/{\nu})$$^{d}$ & 1.9~$\times$~10$^{-1}$             \\ 
&  &                     \\
 ${7}$  &$\Gamma$            & 2.27$_{-0.07}^{+0.07}$                          \\
&   $N_{\rm H}$                  & 0.10 $_{-0.03}^{+0.04}$$\times$10$^{22}$~cm$^{-2}$    \\
&     A$_{\rm Fe}$                & 1 A$_{\odot}$ (fixed)                 \\
&    $a$                                & 0.9 (fixed)  \\
&      Inclination                    & 33$^{+10}_{-13}$ degrees (90 $\%$  confidence)            \\
& $r_{\rm in}$                         & ISCO (fixed)\\
& $r_{\rm out}$                        & 400$r_{\rm g}$ (fixed) \\
&  $R_{\rm frac}$                      & 1.4$^{+0.6}_{-0.6}$  (68 $\%$  confidence)   \\
&    $\log\xi_{\rm disk}$          & 1.7$_{-0.2}^{+0.2}$~erg~cm~s$^{-1}$               \\
 &   N$_{\rm H,abs1}$        & 3.4$_{-2.0}^{+1.9}$ $\times$ 10$^{23}$~cm$^{-2}$   \\
  &    $\log\xi_{\rm abs1}$          & 3.3$_{-0.1}^{+0.3}$~erg~cm~s$^{-1}$               \\
 & $v_{\rm turb,abs1}$ &    18,500 km~s$^{-1}$  (fixed)                  \\
  & $f_{\rm c, abs1}$ &    $>$ 0.5 (90 $\%$  confidence)                  \\
   &   $z_{\rm abs1}$                  &  1.22$_{-0.04}^{+0.04}$                                         \\ 
   &   N$_{\rm H,abs2}$        & 2.9$_{-1.6}^{+2.0}$ $\times$ 10$^{23}$~cm$^{-2}$   \\
  &    $\log\xi_{\rm abs2}$          & 3.2$_{-0.1}^{+0.2}$~erg~cm~s$^{-1}$ \\
& $v_{\rm turb,abs2}$ &  30,000 km~s$^{-1}$ (fixed)                   \\
 &   $z_{\rm abs2}$                  & 0.63$_{-0.15}^{+0.08}$  \\
&   $\chi^2/{\nu}$               & 214.7/196             \\
&  $P(\chi^2/{\nu})$$^{d}$ & 1.7~$\times$~10$^{-1}$             \\ 
&  &                    \\
\hline 
\end{tabular}
\end{center}
\noindent
${}^{a}$ Model 6  consists of a power law, neutral absorption at the source, two outflowing ionized absorbers at the source, scattering of the low velocity outflowing component (abs1), and neutral Compton reflection with self-consistent Fe and Ni lines (PEXMON). The normalizations of the direct(reflection) component is the photon flux at 1 keV (photons~keV$^{1}$~cm$^{-2}$~s$^{-1}$) of the power law(reflection) component only and in the earth frame.
Model 7  consists of a power law, neutral absorption at the source, two outflowing ionized absorbers at the source, scattering of the low velocity outflowing component (abs1),
and relativistic reflection off an ionized disk.  Parameters of model 7 include the spin of the black hole ($a$), the inner and outer radii of the accretion disk $r_{\rm in}$  and $r_{\rm out}$, respectively. For models 6 and 7   
$R_{\rm frac}$ is defined as the ratio of the flux reflected by the disk to the flux of the direct power-law component in the rest-frame energy band of 20$-$40~keV.
Parameters listed as fixed are held fixed to the listed value during the fit.
All model fits include the Galactic absorption toward the source (Dickey \& Lockman 1990).\\
${}^{b}$All absorption-line parameters are calculated for the rest frame.\\
${}^{c}$Spectral fits were performed using the $\chi^{2}$ statistic and all errors are for 90\% confidence unless mentioned otherwise. \\
${}^{d}$$P(\chi^2/{\nu})$ is the probability of exceeding $\chi^{2}$ for ${\nu}$ degrees of freedom
if the model is correct.\\
\end{table}


\clearpage
\begin{table}
\caption{Soft (0.2-2~keV) and Hard (2-10~keV) Fluxes and Luminosities of Quasar \hs\ } 
\scriptsize
\begin{center}
\begin{tabular}{lccccc}
\hline\hline
 Observation      & Observatory    &  $f_{\rm 0.2-2~keV}$$\tablenotemark{a}$         & $f_{\rm 2-10~keV}$$\tablenotemark{a}$      & $L_{\rm 0.2-2~keV}$$\tablenotemark{a}$ & $L_{\rm 2-10~keV}$$\tablenotemark{a}$  \\
     Date              &                                                    &  ($\times$~10$^{-13}$ erg~s$^{-1}$~cm$^{-2}$)                           &   ($\times$~10$^{-13}$erg~s$^{-1}$~cm$^{-2}$)            &  ($\times$~10$^{45}$erg~s$^{-1}$) &  ($\times$~10$^{45}$erg~s$^{-1}$)   \\
\hline
 &  &  &   & &\\
2002 January  30  & {\sl Chandra}                            &  4.7$^{+0.2}_{-0.7}$              & 8.0$^{+1.5}_{-1.3}$         & 4.9$^{+0.4}_{-1.0}$             & 7.6$^{+0.5}_{-0.7}$          \\ 
2013 December 16 & {\sl Chandra}                          &  4.0$^{+0.3}_{-0.4}$              &  3.3$^{+0.1}_{-0.2}$        & 6.1$^{+0.5}_{-0.5}$             & 4.8$^{+0.4}_{-0.4}$  \\ 
2014 October 4       & {\sl XMM-Newton}                  &  6.4$^{+0.3}_{-0.3}$              &  3.1$^{+0.1}_{-0.1}$        & 13.9$^{+0.5}_{-0.7}$            & 4.7$^{+0.2}_{-0.3}$                  \\
&  &  &   & &\\
\hline 
\end{tabular}
\end{center}
${}^{a}${These are unabsorbed fluxes and luminosities for the combined images. The errors are estimated at 90\% confidence. The luminosities are not corrected for the lensing magnification which we estimate to be $\mu_{A+B+C+D}$ $\sim$ 103.}\\

\end{table}

\clearpage 
\begin{table}
\caption{Hydrogen Column Densities, Outflow Velocities, Mass-Outflow Rates and Efficiencies of Outflow in \hs}
\scriptsize
\begin{center}
\begin{tabular}{llccccc}
\hline\hline
Observation & Lines & Absorber & $N_{\rm H}$ & $v_{\rm abs}$  &  $\dot{M}$ & $\epsilon_{\rm k}$  \\
 Date      &&          &           (cm$^{-2})$  & $(c)$ & (M$_{\odot}~yr^{-1}$)  &       \\
\hline
& & &    &    & & \\
2013 December 16 &  \SiXIV\ $1s-2p$, \FeXXV\ $1s^2-1s2p$ & abs1              & $2.9_{-0.4}^{+1.6}$ $\times$ 10$^{23}$    & $ 0.10_{-0.01}^{+0.01}$   & $1.5_{-0.9}^{+1.2}$  &$0.24_{-0.15}^{+0.20}$   \\
2014 October 4       & \FeXXV\ $1s^2-1s2p$     & abs2                                    & $3.4_{-2.0}^{+1.9}$ $\times$ 10$^{23}$ &$0.12_{-0.01}^{+0.02}$& $2.1_{-1.4}^{+2.0} $        & $0.5_{-0.3}^{+0.5}$ \\
2014 October 4       & \FeXXV\  $1s^2-1s2p$   & abs3                                     & $2.9_{-1.6}^{+2.0}$ $\times$ 10$^{23}$ &$0.41_{-0.04}^{+0.07}$& $3.4_{-2.1}^{+2.7}$         & $9_{-6}^{+8}$ \\
& & &    &    & & \\
\hline
\end{tabular}
\end{center}
\tablecomments{The parameters assumed in the estimates of the kinematics of the absorbers are taken
from the best-fit values obtained from model 5 of Table 4 and model 7 of Table 5. 
We adopt a mean value for the bolometric luminosity of $L_{\rm Bol}$ = 1.8 $\pm$ 1.3 $\times$~10$^{45}$erg~s$^{-1}$,
based on the X-ray, UV, and optical bolometric corrections. 
The efficiency of the outflow is defined as $\epsilon_{\rm k} = (1/2)\dot{M}v^{2}/L_{\rm bol}$.\\}
\end{table}




\clearpage

 \begin{figure}
  \includegraphics[width=16cm]{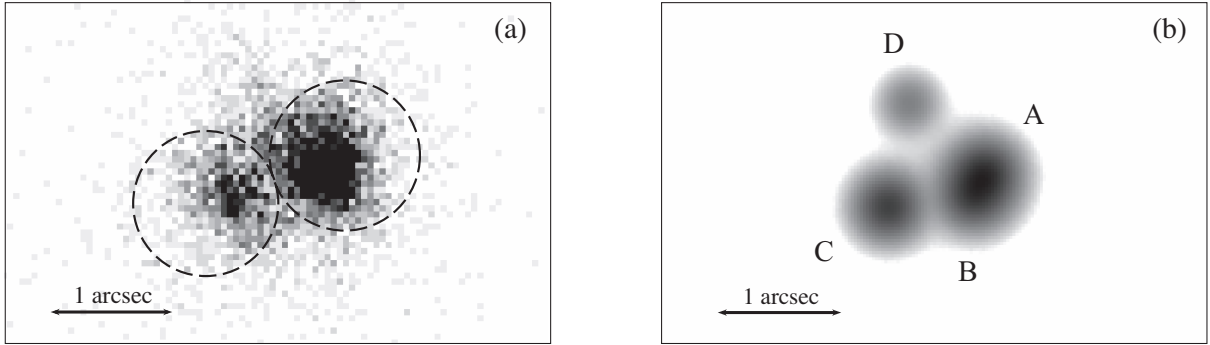}
        \centering
\caption[]{
Images of the 2013 December 16 {\sl Chandra} observation of quasar \hs. (a) Raw 0.2-10~keV image binned with a bin size of 0\sarc1 on a side. The dashed circles show the regions used for spectral extraction. (b) Best fit PSF model to the same observation of \hs.
In all panels north is up and east is to the left.}
\label{fig:images}
\end{figure}

 \begin{figure}
   \includegraphics[width=16cm]{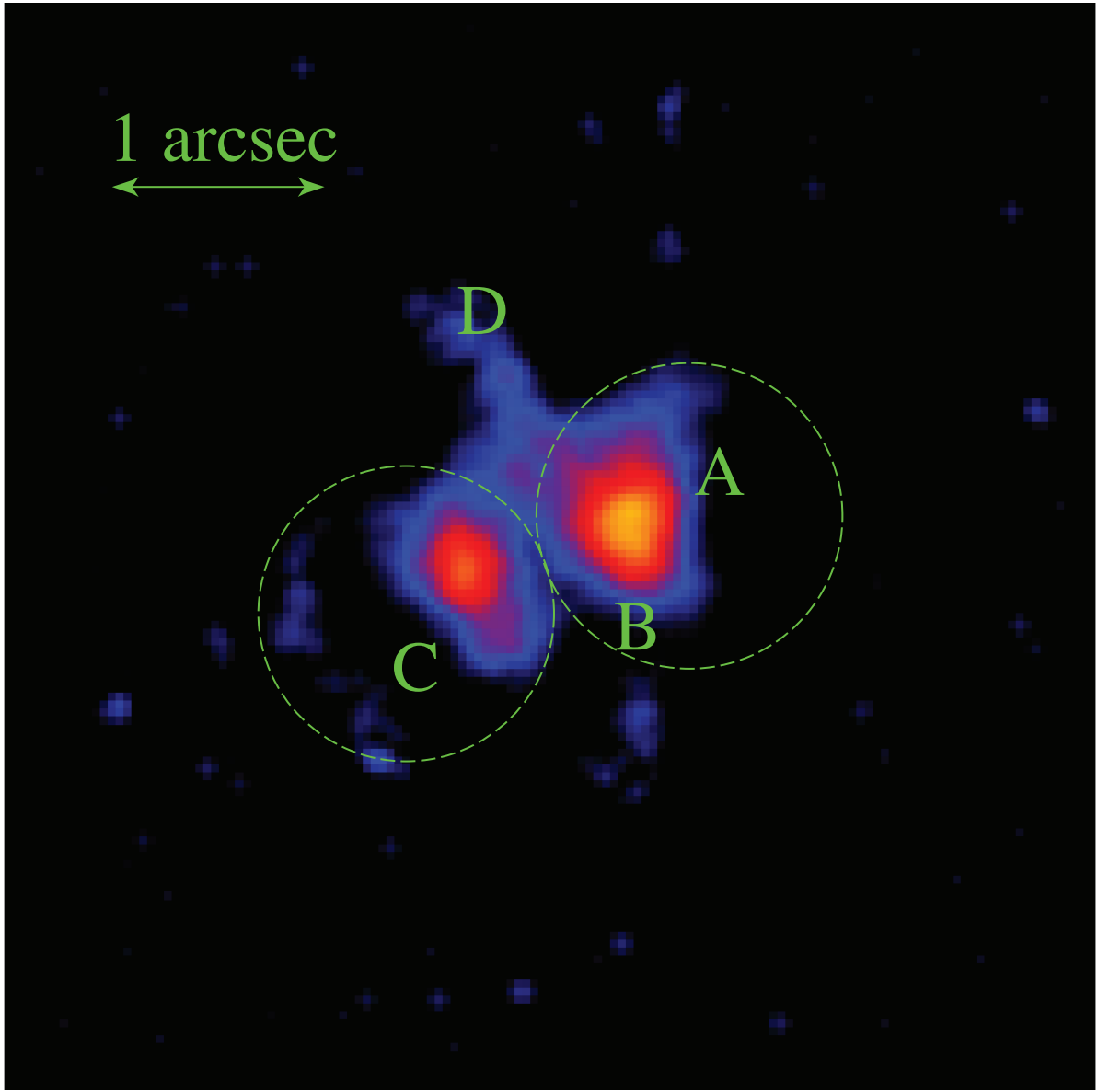}
        \centering
\caption[]{ 
Deconvolved image (0.2$-$10~keV band) of the 2013 observation of \hs. Images A and B are not resolved and image D is not reconstructed likely due to the low S/N of the {\sl Chandra} observation.  The dashed circles show the regions used for spectral extraction. North is up and east is to the left.}
\label{fig:images}
\end{figure}

 \begin{figure}
   \includegraphics[width=12cm]{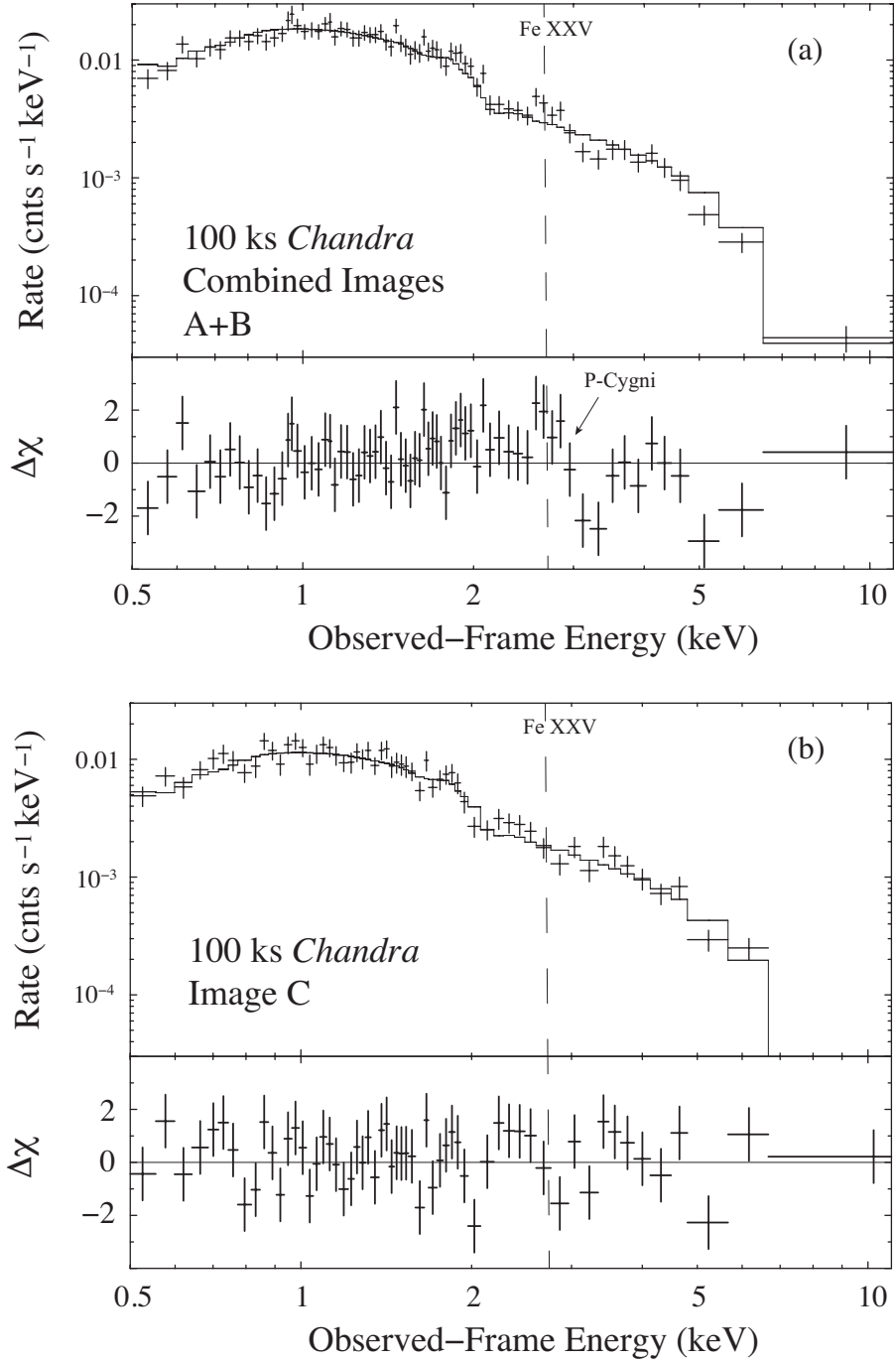}
        \centering
\caption[]{(a) The top panel shows the \chandra\ spectrum of images A+B of
\hs\ fit with Galactic absorption and a 
power-law model. 
The lower panel shows the residuals of the fit in units of 1$\sigma$ deviations.
Several absorption and emission features within the observed-frame ranges of 0.8--1.0~keV and 2--7.0~keV are noticeable in the residuals plot (see Table 3). 
(b) The top panel shows the \chandra\ spectrum of image C of
\hs\ fit with Galactic absorption and a 
power-law model.
The lower panel shows the residuals of the fit in units of 1$\sigma$ deviations.
The spectrum of image C appears different form that of image (A+B) with several weak emission features detected within the observed-frame range of 2--4~keV.}
\label{fig:images}
\end{figure}


\begin{figure}[htb]
      \includegraphics[width=16.cm]{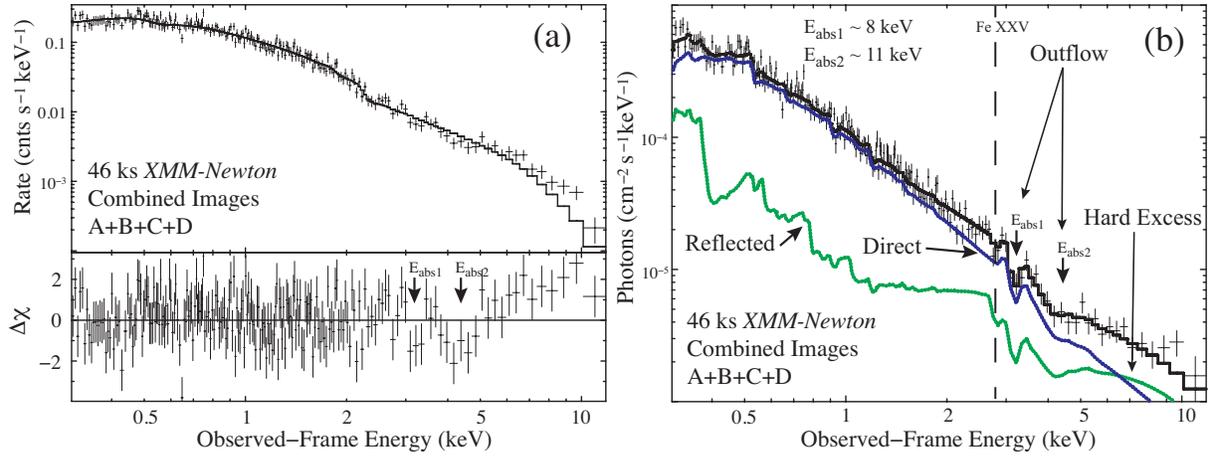}
        \centering
        \vspace{0.in}
\caption[]{
(a) Observed 46~ks {\it XMM-Newton} spectrum of combined images A+B+C+D of \hs\ fit with Galactic absorption and a power-law model. 
Notice significant hard X-ray excess.
(b) shows the data shown in panel (a) overplotted with the unfolded best-fit model (model 7 of Table 5)
comprised of a photoionizaton outflow (XSTAR), a direct component (blue curve) and a reflection component (red curve).
The direct and reflected components are assumed to be absorbed by the outflowing wind.
The arrows indicate the best-fit energies of the absorption lines and the reflection and direct components. 
}
\label{fig:images}
\end{figure}


 \begin{figure}
   \includegraphics[width=16cm]{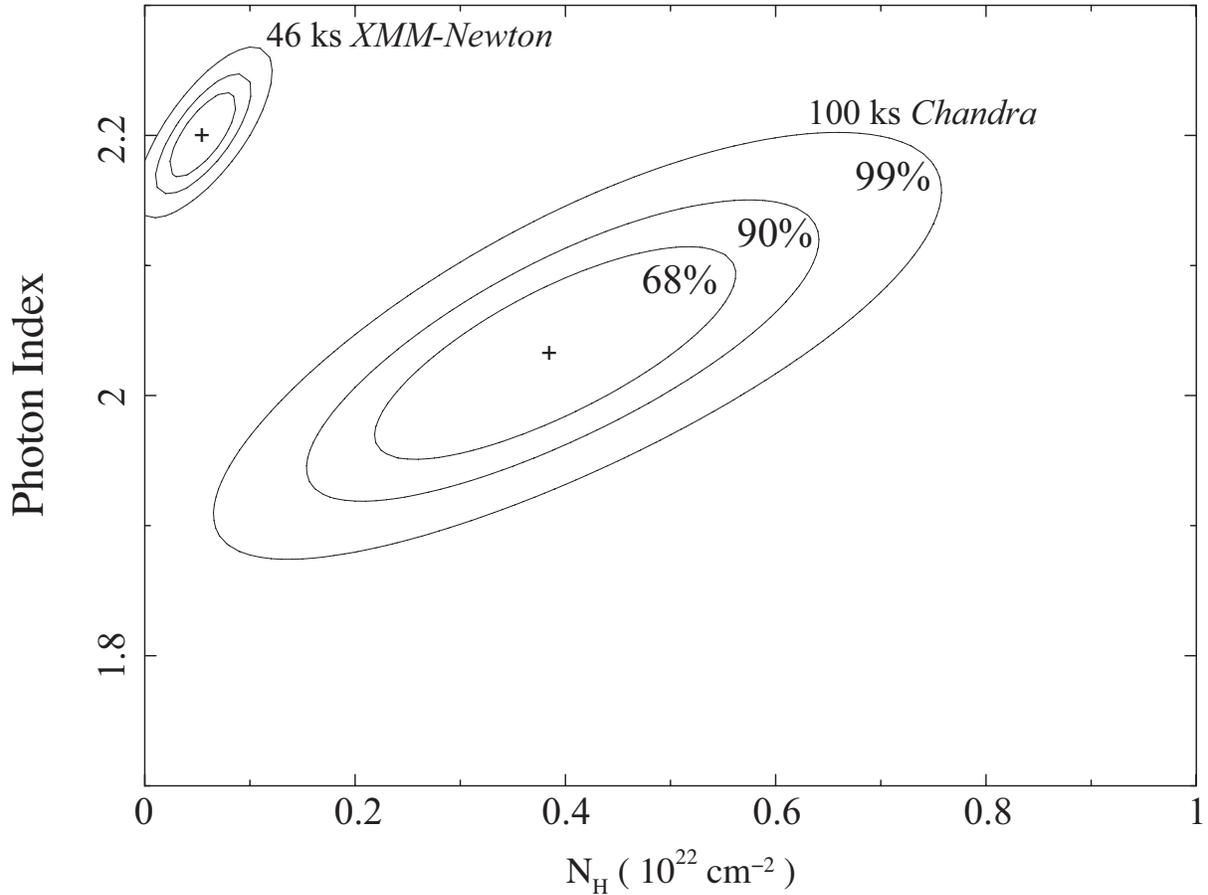}
        \centering
\caption[]{68\%, 90\%, and 99\%  $\chi^2$ confidence contours of
intrinsic  N$_{\rm H}$ versus photon index obtained in the fits to the {\it Chandra} spectrum of image A+B and the {\sl XMM-Newton} spectrum of the combined images.
Both spectral fits used model 3 of Table 3. }
 \label{fig:images}
\end{figure}

 \begin{figure}
  \includegraphics[width=16cm]{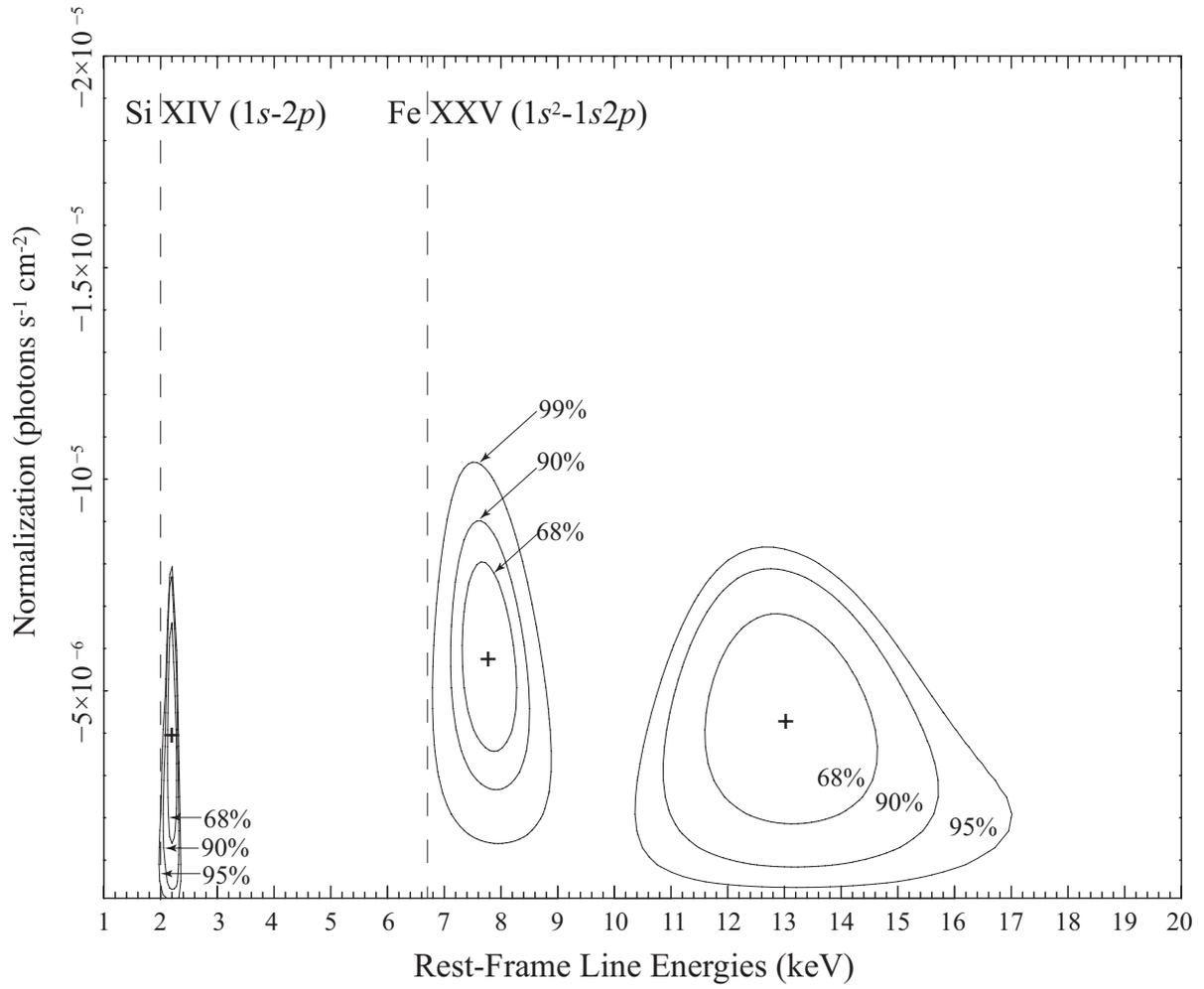}
        \centering
\caption[]{$\chi^2$ confidence contours between the normalizations of the absorption lines based on fits using model 2 of Table 3 to the {\it Chandra} spectrum of image A+B.
The 99\% confidence contours, of $E_{\rm abs1}$ and $E_{\rm abs3}$ are erratic and not closed at the
99\% level and are therefore not displayed.}
\label{fig:images}
\end{figure}

 \begin{figure}
   \includegraphics[width=16cm]{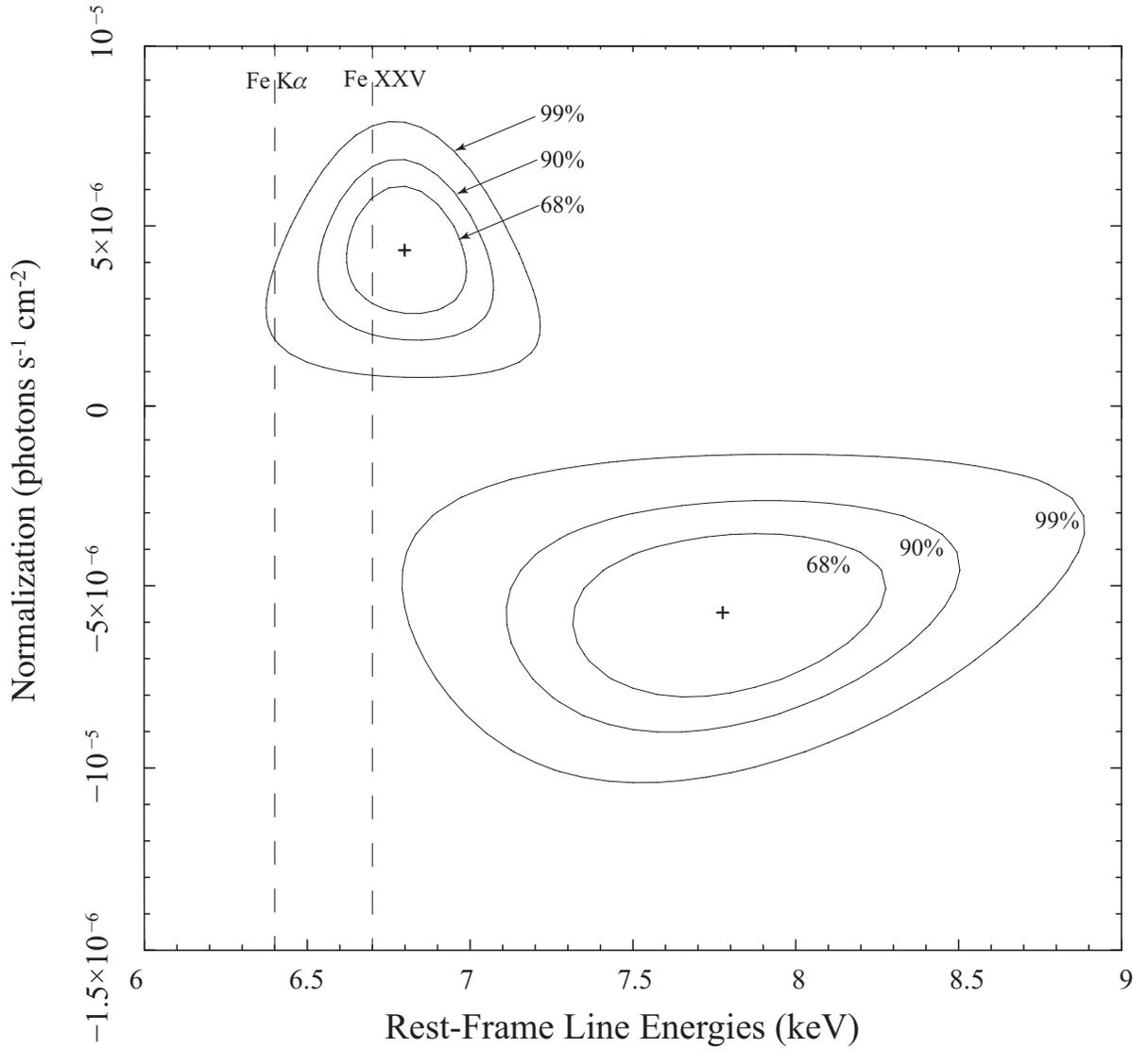}
        \centering
\caption[]{$\chi^2$ confidence contours 
between the normalizations of the P-Cygni line detected in the {\it Chandra} spectrum of image A.}
\label{fig:images}
\end{figure}

 \begin{figure}
   \includegraphics[width=16cm]{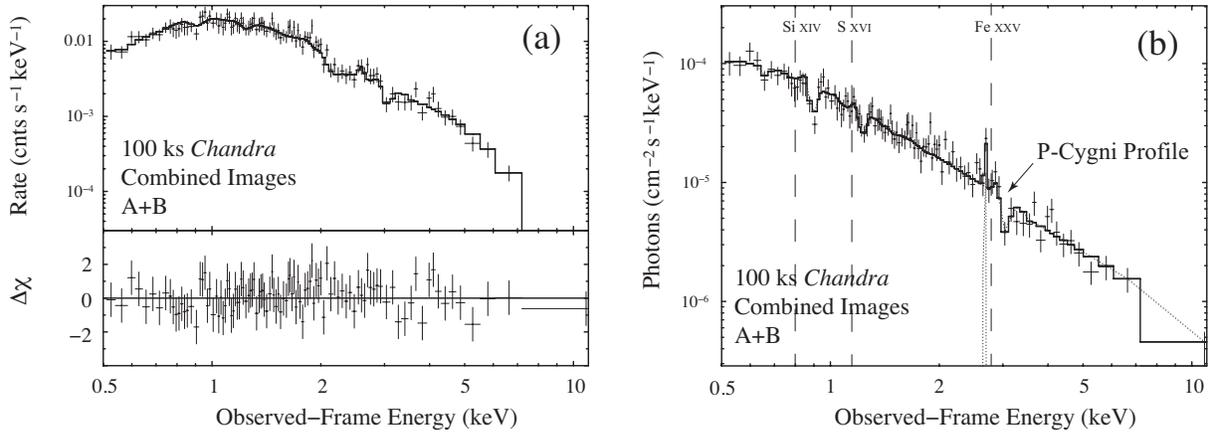}
        \centering
\caption[]{ (a) Observed 100~ks {\it Chandra} spectrum of combined images A+B of \hs\ fit with model 5 of Table 4 that includes a P-Cygni component. 
(b) The unfolded best-fit model to the fit shown in panel (a). The vertical dashed lines indicate the observed-frame energies of the Si XIV, S XVI, and Fe XXV resonance spectral lines. The inferred common outflow velocity of the absorber is v = 0.1$c$.
}
\label{fig:images}
\end{figure}

 \begin{figure}
   \includegraphics[width=16cm]{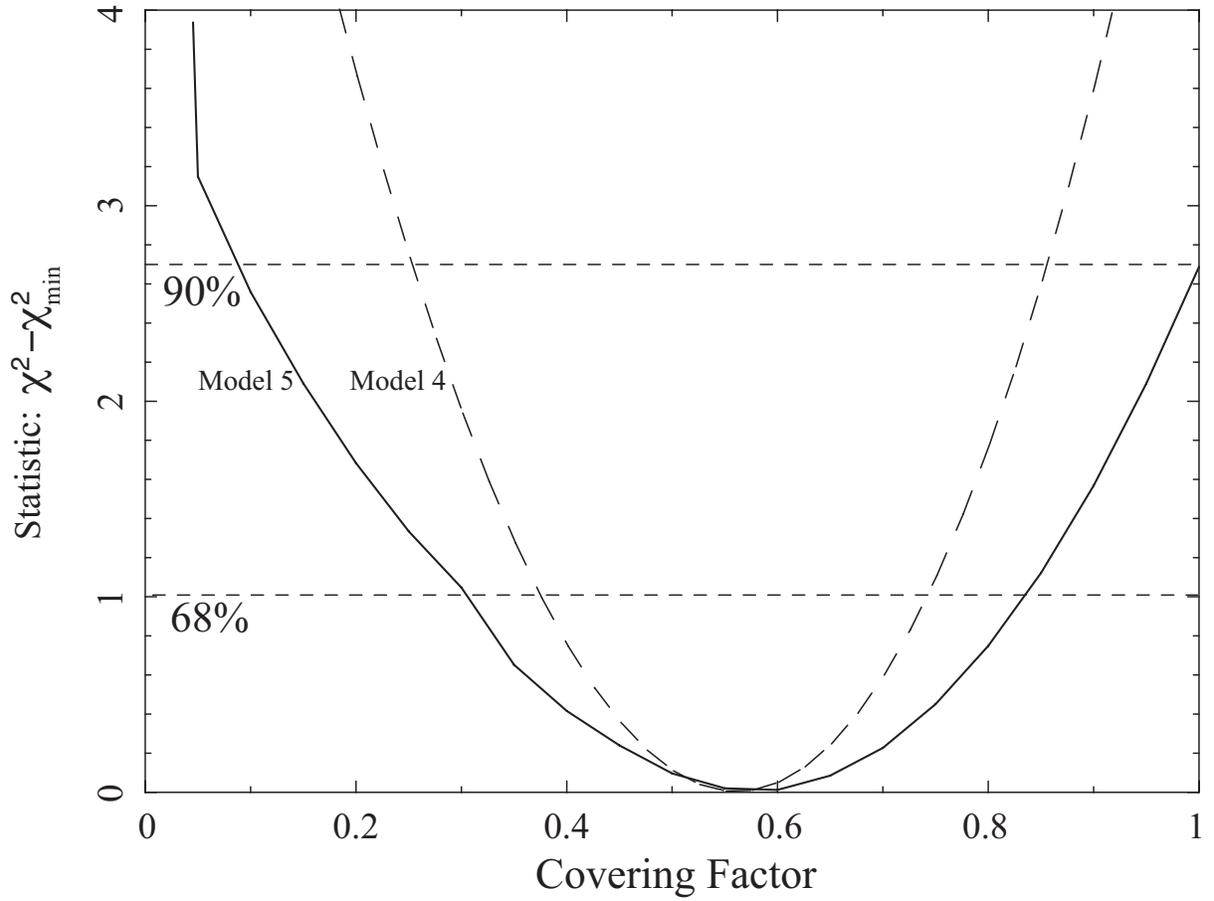}
        \centering
\caption[]{$\chi^{2}$ confidence contours of the covering factor of the wind for models 4 and 5 of Table 4. The best-fit values of the covering factor are 
$f_{\rm c}$ = 0.57$_{-0.18}^{+0.18}$ (68$\%$)  assuming model 4 and  $f_{\rm c}$ = 0.6$_{-0.3}^{+0.2}$ (68$\%$) for model 5.
}
\label{fig:images}
\end{figure}

 \begin{figure}
   \includegraphics[width=16cm]{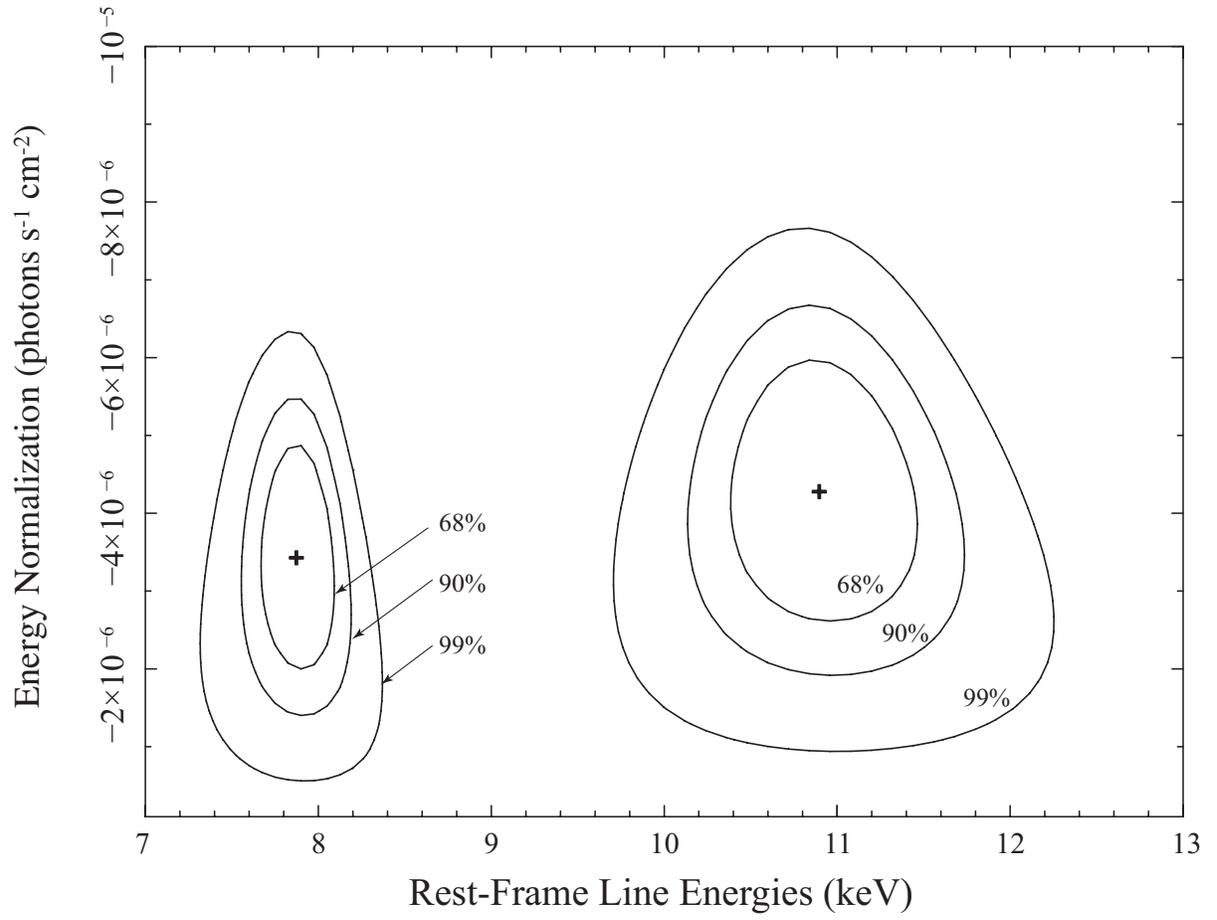}
        \centering
\caption[]{$\chi^{2}$ confidence contours between the normalizations of the absorption lines and their rest-frame energies based on fits using model 2 of Table 3 to the {\it XMM-Newton} spectrum of image A+B+C+D.
}
\label{fig:images}
\end{figure}

 \begin{figure}
   \includegraphics[width=16cm]{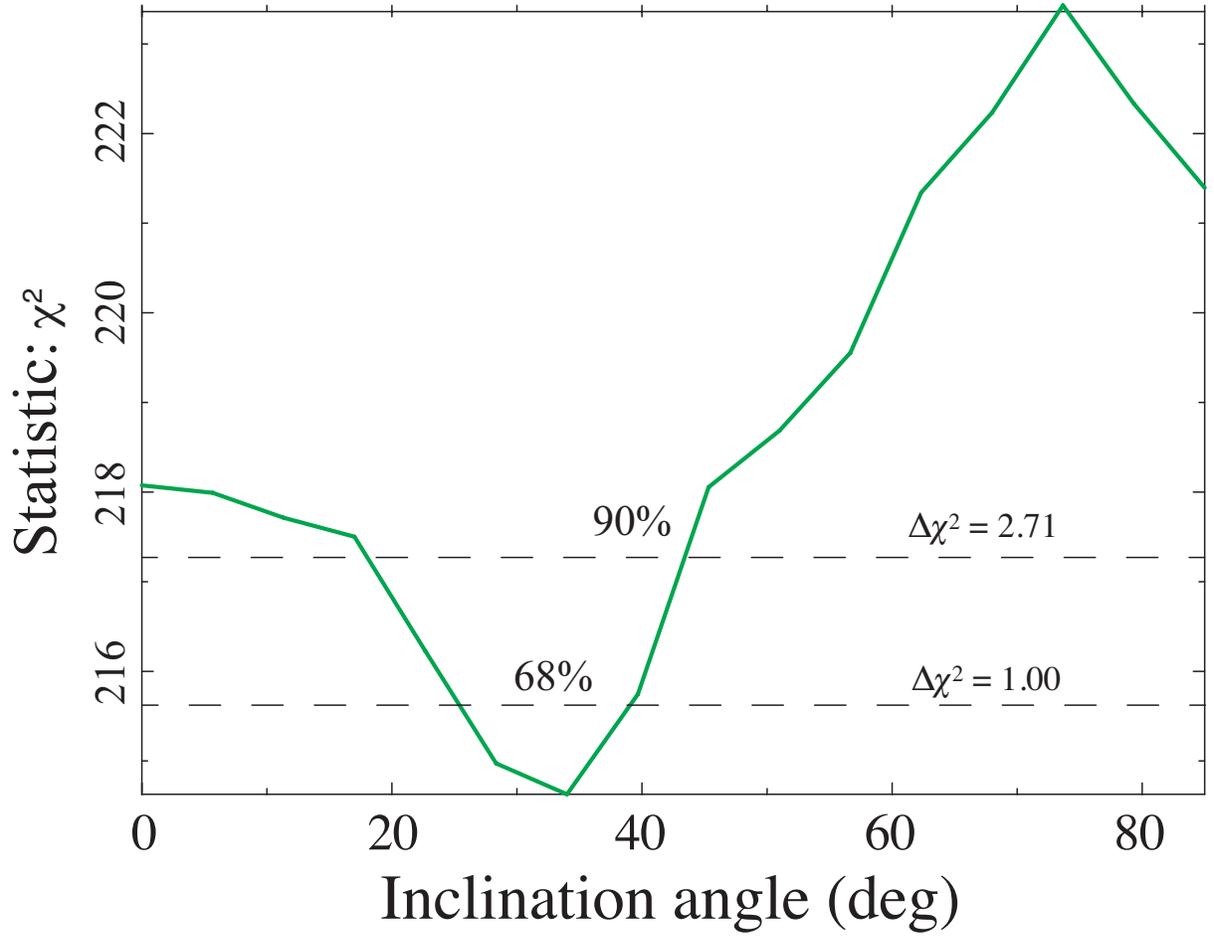}
        \centering
\caption[]{Fits to the {\sl XMM-Newton} spectrum of \hs\ constrain the inclination angle of the accretion disk (model 7 of Table 5). 
The $\chi^{2}$ confidence plot indicates an inclination angle of $<$ 45$^{\circ}$ (90\% confidence) consistent with models that posit NALQSOs as objects observed at relative low inclination angles. 
}
\label{fig:images}
\end{figure}

 \begin{figure}
  \includegraphics[width=16cm]{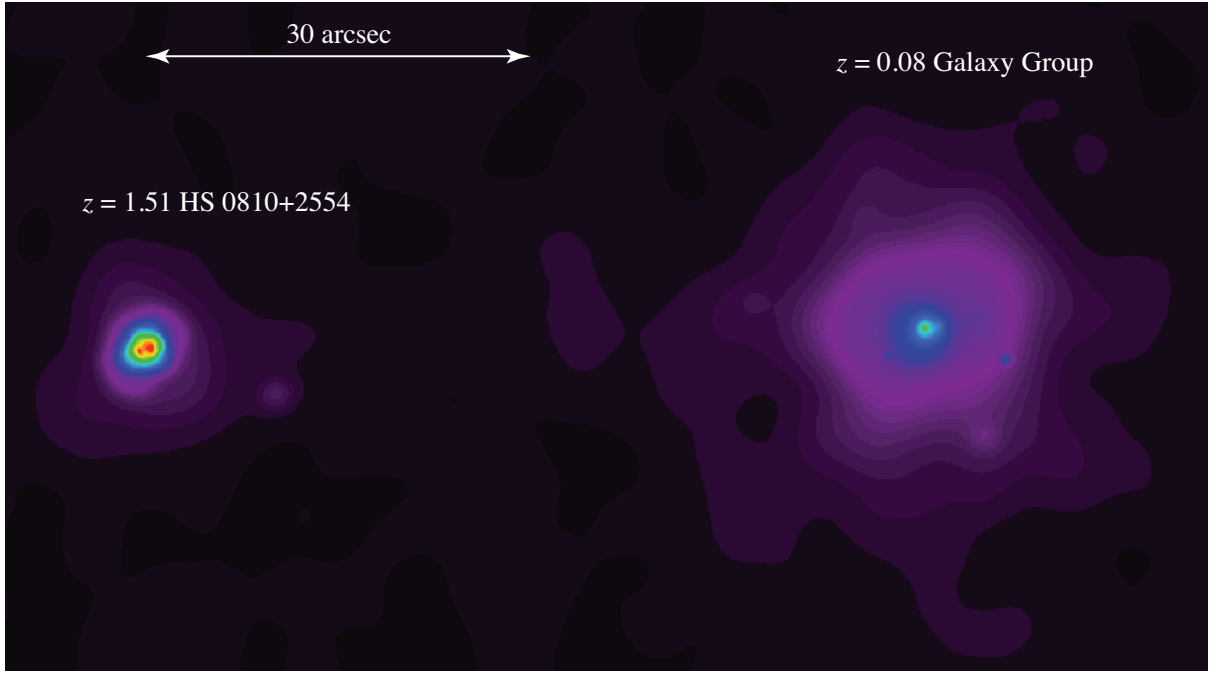}
        \centering
\caption[]{Adaptively smoothed image of the 100~ks \chandra\ observation of \hs. 
Our spectral analysis confirms our previous result that was based on a 5~ks exposure (Chartas et al. 2014) that the extended emission 58 \arcsec\ west of \hs\ is produced by a galaxy group at $z \sim 0.08$.}
\label{fig:images}
\end{figure}

 \begin{figure}
   \includegraphics[width=16cm]{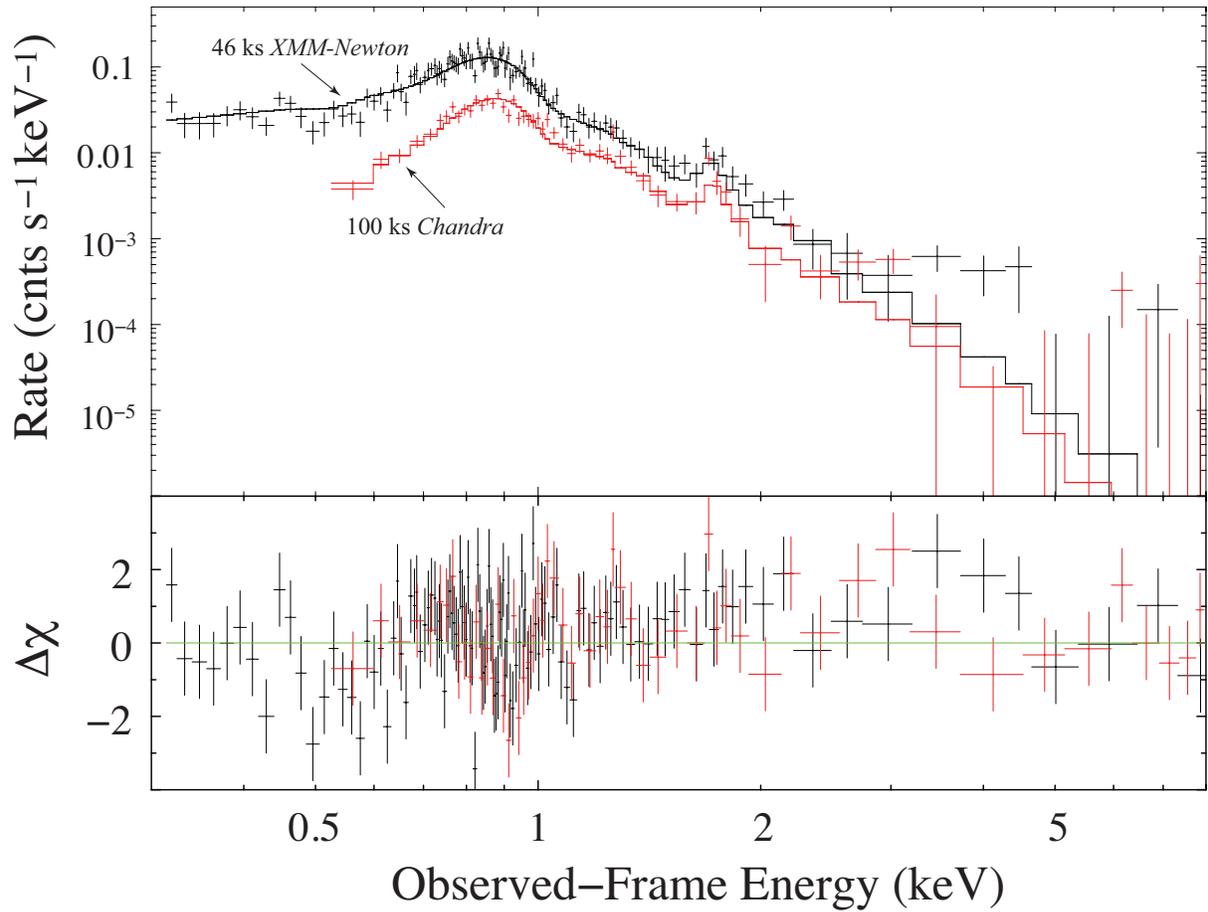}
        \centering
\caption[]{The {\sl Chandra} and {\sl XMM-Newton} X-ray spectra of the galaxy group centered 58\arcsec\ west of  \hs\ along with the 
best-fit thermal {\it mekal} model (joint fit). The model is consistent
with a plasma temperature of $T_{\rm e}$ $\sim$ 0.8 keV, abundance of  $A$ $\sim$ 0.53 solar and a redshift of $z$ $\sim$ 0.08.  (Lower panel) Residuals in units of standard deviations with error bars of size 1$\sigma$. }
\label{fig:images}
\end{figure}

\end{document}